\documentclass[aps,twocolumn,groupedaddress,amsmath,amssymb,nofootinbib]{revtex4-2}
\usepackage[normalem]{ulem}
\usepackage{graphicx}
\usepackage{xcolor}
\usepackage{ulem}
\begin{document}

\title{\textcolor{black}{Thermodynamics} of non-linearly charged \textcolor{black}{Anti-de Sitter} black holes in four-dimensional Critical Gravity}

\author{Abigail \'Alvarez}
\email{abialvarez-at-uv.mx}
\affiliation{Universidad de Xalapa, Carretera Xalapa-Veracruz No. 341, 91190, Xalapa, Ver., M\'exico}
\affiliation{ Facultad F\'{\i}sica, Universidad Veracruzana, 91000, Xalapa, Ver., M\'exico.}

\author{Mois\'es Bravo-Gaete}
\email{mbravo-at-ucm.cl}
\affiliation{Facultad de Ciencias B\'asicas, Universidad Cat\'olica del Maule, Casilla 617, Talca, Chile.}

\author{Mar\'ia Montserrat Ju\'arez-Aubry}
\email{mjuarez-at-astate.edu}
\affiliation{Arkansas State University Campus Queretaro, Carretera estatal $\#100$ km. 17.5, Municipio Col\'on, 76270, {Quer\'etaro}, M\'exico}

\author{Gerardo Vel\'azquez Rodr\'iguez}
\email{gvelazquez-at-astate.edu}
\affiliation{Arkansas State University Campus Queretaro, Carretera estatal $\#100$ km. 17.5, Municipio Col\'on, 76270, {Quer\'etaro}, M\'exico}

\begin{abstract}
In this work, we provide new examples of Anti-de Sitter black holes with a planar base manifold in four-dimensional Critical Gravity by considering nonlinear electrodynamics as a matter source. \textcolor{black}{We find a general solution characterized by the presence of only one integration constant   where, for a suitable \textcolor{black}{choice} of coupling constants, we can show the existence of one or more horizons. Additionally,}  we compute its \textcolor{black}{nonzero} thermodynamical quantities through a variety of techniques, testing the validity of the first law of thermodynamics as well as a Smarr formula. Finally, we analyze the local thermodynamical stability of the solutions. \textcolor{black}{To our knowledge, these charged configurations are the first example with Critical Gravity where their thermodynamical quantities are not zero.}
\end{abstract}

\maketitle
%\tableofcontents
\newpage

%%%%%%%%%%%%%%%%%%%%%%%%%%%%%%%%%%%%%%%%%%%%%%%%%%%%%%%%%%%%%%%%%%%%%%%%%%%%%%%%%%
\section{Introduction}
%%%%%%%%%%%%%%%%%%%%%%%%%%%%%%%%%%%%%%%%%%%%%%%%%%%%%%%%%%%%%%%%%%%%%%%%%%%%%%%%%%

Since its introduction in the late nineties, the idea of the \textcolor{black}{Anti-de Sitter / Conformal Field Theory (AdS/CFT)} correspondence \cite{Maldacena:1997re} has gained momentum in different areas due to its potential to shed light on phenomena that range from superconductivity to quantum computing. This correspondence has also generated the need for the study of theories other than General Relativity due to two main reasons: first, the necessity to have theories that exhibit desired symmetries and properties that match the non-relativistic systems in the context of the correspondence and, secondly, the possibility that these \emph{enhanced} theories may support a variety of thermodynamically rich AdS black holes, whose holographic role will be to introduce the non-relativistic behavior at finite temperature.
 %The reason behind this last need is that the holographic role of black holes is to introduce the non-relativistic behavior at finite temperature.
%The existence of black holes is of particular interest since it is through them that thermodynamics is translated from the gravitational theory into the gauge theory.

In this context, quadratic curvature gravities have been very successful in providing a plethora of new black hole configurations \cite{Cai:2009ac,Ayon-Beato:2009rgu,AyonBeato:2010tm,Alishahiha:2011yb,Pravda:2020zno,Svarc:2018coe}. A notable example within these theories in 2+1 dimensions is New Massive Gravity \cite{Bergshoeff:2009hq}, a parity-even, renormalisable theory that, at the linearized level, is equivalent to the unitary Fierz-Pauli theory for free massive spin-2 gravitons, where AdS solutions have been previously found \cite{Bergshoeff:2009aq,Oliva:2009ip}. A four-dimensional analogue to this theory is Critical Gravity (CG) \cite{Lu:2011zk}, { which is a ghost-free, renormalizable theory of gravity with quadratic corrections in the curvature\footnote{{{In general, a theory with quadratic corrections in curvature will propagate massive scalar and spin-2 ghost fields. In CG, the relation between the coupling constants and the cosmological constant leads to a theory in which the scalar field is zero and the spin-2 field becomes massless and with vanishing energy.}} }.}
A particularity of CG is that its vacuum admits an AdS black hole solution given by
\begin{eqnarray}\label{eq:vacuum_sol}
ds^2&=&\displaystyle{-\frac{r^2}{l^2} \left(1-\frac{M l^3}{r^3}\right) dt^2 + \frac{l^2}{r^2} \frac {dr^2}{\left(1-\frac{M l^3}{r^3}\right)}} \nonumber \\
&&+\frac{r^2}{l^2} \left(dx_1^2 +dx_2^2\right),
\end{eqnarray}
where $M$ is an integration constant. However, as emphasized by the authors in \cite{Lu:2011zk}, this solution is massless and has a vanishing entropy. {This is also highlighted in \cite{Anastasiou:2017rjf, Anastasiou:2021tlv} in four and six dimensions, where the entropy, as well as global conserved charges of their black holes solutions, vanish identically.}

Because {of} the interest of obtaining black hole solutions that can be framed in the context of the \textcolor{black}{gauge/gravity} correspondence, in this work we aim to find new AdS black hole configurations that can be supported by CG and exhibit non-vanishing thermodynamical properties. To achieve this, a nonlinear electrodynamics (NLE) source
is employed \cite{Plebanski:1968}. \textcolor{black}{
The origin of NLE dates to $1912$, the year in which Mie G.  explored this formalism  for the first time \cite{Mie:1912}. Some years later, in the thirties, and motivated in part to avoid the well-known singularity of the field of a point particle,  Born and Infeld \cite{Born:1934,Born.Infeld:1934,Born.Infeld:1935} proposed new research %with high impact
, giving rise to the Born-Infeld (BI) theory \cite{Weiss:1937jtb,Infeld:1936,Infeld:1937}.   Given the complexity to carry out an extension of BI towards solutions of nonlinear equations, this formalism had a stage of stagnation for almost three decades. Nevertheless, around the sixties, J. F. Pleb\'anski presented an outstanding work for NLE in a medium, in which the theory is developed through an antisymmetric conjugate tensor $P^{\mu\nu}$ (known as of Pleb\'anski tensor) and a structure-function $\mathcal{H}=\mathcal{H}(P,Q)$, where $ P$ and $Q$ are the invariants that are formed with the antisymmetric conjugate Pleb\'anski tensor. Here, the structure-function $\mathcal{H}(P,Q)$ is associated with the Lagrangian function $\mathcal{L}(F,G)$, that depends on the invariants quadratics constructed from the Maxwell tensor $F_{\mu\nu}$, which can be determined by  a Legendre transformation.  At the end of the eighties, H. Salazar, A. Garc\'ia, and J. Pleb\'anski found solutions for the equations of NLE coupled to gravity using the formalism \cite{Plebanski:1968}, in which the BI theory appears as a special case \cite{Salazar:1987}. Some of the benefits provided by the Pleb\'anski formalism is the ability to obtain regular black hole solutions \cite{Ayon-Beato:1998hmi,Ayon-Beato:1999qin,Ayon-Beato:1999kuh,Ayon-Beato:2000mjt,Ayon-Beato:2004ywd}, Lifshitz black hole configurations that exist for any value of the dynamic exponent $z > 1$  \cite{Alvarez:2014pra}, and recently an exact solution of a massive, electrically and magnetically charged, rotating stationary black hole has been found \cite{Garcia-Diaz:2021bao,Diaz:2022roz,Ayon-Beato:2022dwg}. In General Relativity,  NLE has been a valuable tool in order to build exact black hole configurations, some of which exhibit non-standard asymptotic behavior in Einstein's gravity or in its generalizations, as can be verified in \cite{Hassaine:2007py, Hassaine:2008pw, Maeda:2008ha,HabibMazharimousavi:2013duq, Diaz-Alonso:2011riv, Roychowdhury:2012vj,Jing:2011vz}.  It is relevant to mention that, in these works, charged black hole solutions that come from nonlinear theories possess  interesting thermodynamic properties
\cite{Gonzalez:2009nn,Dehghani:2013mba,Bokulic:2021dtz,Liu:2014dva, Bravo-Gaete:2020ftn, Bravo-Gaete:2015xea}. All the above shows NLE as an interesting and motivating study field that we aim to explore with the addition of CG}.
%Some examples of NLE's relevant role in exact configurations include the following: in some cases it regularizes the singularities of spacetime, and {in} other cases it allows non-standard asymptotic behaviors in Einstein gravity, or some of its generalizations and also endows them with good thermodynamic properties that make them interesting for its analysis {\cite{Ayon-Beato:1998hmi,Ayon-Beato:1999qin,Ayon-Beato:1999kuh,Ayon-Beato:2000mjt,Ayon-Beato:2004ywd,Hassaine:2007py,
%Hassaine:2008pw,Maeda:2008ha,Gonzalez:2009nn,HabibMazharimousavi:2013duq,Diaz-Alonso:2011riv,Roychowdhury:2012vj,
%Jing:2011vz,Dehghani:2013mba,Alvarez:2014pra,Bravo-Gaete:2021kgt}}. Previous attempts to explore the nature of the laws of thermodynamics in the presence of non-linear electrodynamics include a generalization of the first law in presence of a nonlinear magnetic field, using the covariant phase space formalism \cite{Bokulic:2021dtz}.
As such, our action of interest will be given by:

%In the interest of obtaining black hole solutions that can be framed in the context of the AdS/CFT correspondence, in this work we aim to find new AdS black hole configurations that can be supported by Critical Gravity and exhibit non-vanishing thermodynamical properties. To achieve this, we will introduce nonlinear electrodynamics (NLE) \cite{Plebanski:1968} as a way to enrich the action of Critical Gravity. As such, our action of interest will be given by:

%With this in mind, the rest of the paper will be organized as follows: In the lines below we will present our action, as well as the equations of motion, in section II we show our general solution, in section III we analyze our general solution and show the particular types of solutions that we can have and characterize them in terms of their maximum number of horizons; in section IV we calculate the thermodynamical quantities of these solutions, we verify that they satisfy the first law of thermodynamics as well as a Smarr formula and, lastly, we study their stability. Finally, in section V, we present our conclusions.

\begin{eqnarray}\label{eq:Squad}
S[g_{\mu \nu},A_\mu, P^{\mu\nu}]
=\int{d}^4x\sqrt{-g}(\mathcal{L}_{CG} + \mathcal{L}_{NLE})\,,
\end{eqnarray}
with
\begin{eqnarray}
\mathcal{L}_{CG} &=& \frac{1}{2\kappa}
\left(R-2\Lambda+\beta_1{R}^2
+\beta_2{R}_{\alpha\beta}{R}^{\alpha\beta} \right)\,,\nonumber\\
\mathcal{L}_{NLE} &=&
-\frac{1}{2}P^{\mu\nu}F_{\mu\nu}+\mathcal{H}(P)\,,\nonumber
\end{eqnarray}
where $\Lambda$ is the cosmological constant. As stated above, CG allows for the massive spin-$0$
field to vanish if the coupling constants $\beta_1$ and $\beta_2$ are restricted
to obey the relations
\begin{eqnarray}\label{relations}
\beta_{2}=-3\,\beta_{1},\qquad \beta_{1}=-\frac{1}{2\Lambda}.
\end{eqnarray}
Moreover, the Lagrangian density $\mathcal{L}_{NLE}$ describes the nonlinear behavior of the
electromagnetic field $A_\mu$ with field strength
$F_{\mu \nu}:=\partial_{\mu} A_{\nu}-\partial_{\nu} A_{\mu}$.
%The tensor $P_{\mu\nu}$ is an antisymmetric tensor where its components in the
%plane space are precisely the electric induction $\vec{D}$ and the magnetic field
%$\vec{H}$.
The introduction of  $P_{\mu\nu}$, which is an antisymmetric secondary field
function of the original field $F_{\mu\nu}$, arises from the need to establish a
{relationship} between standard electromagnetic theory with Maxwell's theory of
continuous media.

The source, described by the structure function $\mathcal{H}(P)$ is, of course, real
and depends on the invariant formed with the conjugated antisymmetric tensor $P^{\mu\nu}$,
which is $P: = \frac{1}{4}P_{\mu\nu}P^{\mu\nu}$, .
In general, the structural function also depends on the other invariant $\mathcal{Q}=-\frac{1}{4}P_{\mu\nu}{}^*P^{\mu\nu}$
where ${}^*$ represents the Hodge dual. Here, this invariant is zero because we are interested
in static configurations.

%Usually, nonlinear electrodynamics is considered plausible if it satisfies the
%correspondence principle, that is, \textbf{if} in the weak field limit ($P<<1$), the non-linear electrodynamics is reduced
%to Maxwell's theory $\mathcal{H}=\mathcal{H}_{lin}= P$.

%Nonlinear electrodynamics generally respects the same Maxwell symmetries, that is,
%in flat space it respects Lorentz and the symmetry of norm $U(1)$, in curved space
%it respects the general covariance and symmetry $U(1)$.
%In flat space, NLE follows the Lorentz symmetry and symmetry of {Gauge \sout{norm}} $U(1)$(Maxwell symmetries), and
%for curved space NLE the general covariance and $U(1)$ symmetry.
%Moreover, while in four dimensions Maxwell
%is also a conformal invariant, nonlinear electrodynamics in general are not conformal
%invariants (although there are cases in which they are, {see for example} reference \cite{Hassaine:2008pw}).

The field equations % of the critical gravity-nonlinear electrodynamic system
that result from the variation of the action (\ref{eq:Squad})
% with respect to the independent variables involved
are

\begin{subequations}\label{eq:EOM}
\begin{eqnarray}
\nabla_{\mu}P^{\mu\nu}=0, \label{eq:Maxwell}
\end{eqnarray}
\begin{eqnarray}
F_{\mu\nu}=\frac{\partial \mathcal{H}}{\partial P} P_{\mu \nu}= \mathcal{H}_P P_{\mu\nu}\,,
%\textrm{where} \,\,
% \mathcal{H}_P\equiv \partial \mathcal{H}/\partial P.
\label{eq:constitutive}
\end{eqnarray}
\begin{eqnarray}
\textcolor{black}{E_{\mu \nu}:=G_{\mu\nu}+\Lambda g_{\mu\nu}+K_{\mu\nu}^{CG}-\kappa T_{\mu\nu}^{NLE}=0,}
\label{eq:Einstein}
\end{eqnarray}
where the tensors $K^{CG}_{\mu\nu}$ and $T^{NLE}_{\mu\nu}$ are defined as follows:
\begin{eqnarray}
K_{\mu\nu}^{CG}&=&
 2\beta_2 \Bigl(R_{\mu\rho}R_{\nu}^{\, \rho}
 -\frac{1}{4}R^{\rho\sigma}R_{\rho\sigma}g_{\mu\nu}\Bigr)\nonumber \\
&+&2 \beta_1 R \Bigl(R_{\mu\nu}-\frac{1}{4} R g_{\mu\nu}\Bigr)
+\beta_2  \Bigl(\Box R_{\mu\nu} + \frac{1}{2} \Box R g_{\mu\nu}\nonumber \\
&-& 2 \nabla_{\rho}\nabla_{\left(\mu\right.}R_{\left.\nu\right)}^{\,  \rho} \Bigr)
+2\beta_1 (g_{\mu\nu}\Box R-\nabla_{\mu}\nabla_{\nu} R), \nonumber  \\
T_{\mu\nu}^{NLE}&=&  {\mathcal{H}_P} P_{\mu\alpha}P_{\nu}^{\, \alpha}
-g_{\mu\nu}(2P\mathcal{H}_P-\mathcal{H})\,,\nonumber
\end{eqnarray}
\end{subequations}
with $\beta_1$ and $\beta_2$ given previously in (\ref{relations}). Note that equation (\ref{eq:Maxwell}) represents the nonlinear version of Maxwell's equations,
while the constitutive relations are encoded in (\ref{eq:constitutive}) and Einstein's equations are given
by (\ref{eq:Einstein}).

These equations of motion (\ref{eq:Maxwell})-(\ref{eq:Einstein}) will lead us to find new AdS black hole {configurations} in {CG} coupled with non-linear electrodynamics in section II. Then, in section III, we will show
%The rest of the paper will be organized as follows:
the analysis and characterization of said solutions in terms of the maximum number of horizons. In section IV, the thermodynamical quantities {and local stability} of the solutions are calculated, and the first law of thermodynamics as well as the Smarr formula are verified. Finally, in section V we present our conclusions and perspectives of this work.

%In section III we will analyze said solutions and characterize them in terms of their maximum number of horizons; in section IV we will calculate the thermodynamical quantities of these solutions, we will verify that they satisfy the first law of thermodynamics as well as a Smarr formula and, lastly, we will study their stability. Finally, in section V, we will present our conclusions.

%%%%%%%%%%%%%%%%%%%%%%%%%%%%%%%%%%%%%%%%%%%%%%%%%%%%%%%%%%%%%%%%%%%%%%%%%%%%%%%%%%%%%%%%%%%%%
\section{Solution to the Equations of motion}
%%%%%%%%%%%%%%%%%%%%%%%%%%%%%%%%%%%%%%%%%%%%%%%%%%%%%%%%%%%%%%%%%%%%%%%%%%%%%%%%%%%%%%%%%%%%

We start the search of new AdS black hole solutions by considering the following asymptotically AdS metric ansatz

%To determine the desired solution we consider the following asymptotically AdS metric ansatz
%for the background
\begin{eqnarray}\label{eq:ansatz}
ds^2=-\frac{r^2}{l^2} f(r) dt^2 + \frac{l^2}{r^2} \frac {dr^2}{f(r)}
+\frac{r^2}{l^2} \left(dx_1^2 +dx_2^2\right),
\end{eqnarray}
\textcolor{black}{where  $t \in \,(-\infty,+\infty), r > 0$, the planar coordinates both are assumed belong to a compact set, this is $0 \leq x_1 \leq \Omega_{x_1}$ and $0 \leq x_2 \leq \Omega_{x_2}$,} and the gravitational potential must satisfy the asymptotic condition $${\lim_{r\rightarrow +\infty}f(r) = 1.}$$
In our case, we propose the structure function $\mathcal{H}$, determining
the nonlinear electrodynamics, to be given by
\begin{eqnarray}\label{eq:H}
\mathcal{H}(P) &=&  \frac{ (\alpha_2^2-3 \alpha_1 \alpha_3)l^2 P}{3\kappa}
{-\frac{2\alpha_1 (-2 P)^{1/4}}{l\kappa}}\nonumber\\
&+&{\frac{\alpha_2\sqrt{-2 P} }
{\kappa}},
\end{eqnarray}
where $\alpha_1$, $\alpha_2$ and $\alpha_3$ are coupling constants.

For the purposes of this article, we consider purely electrical
configurations, such that
{$P_{\mu\nu}=2\delta^t_{[\mu}\delta_{\nu]}^r D(r)$.} If we replace the previous ansatz in the nonlinear Maxwell equation (\ref{eq:Maxwell}) we obtain
\begin{equation}\label{eq:Pmunu}
P_{\mu\nu}=2\delta_{[\mu}^t\delta_{\nu]}^r\frac{M}{r^2}.
\end{equation}
Therefore, the electric invariant $P$ is negative definite, since we only consider purely electrical configurations, which reads
\textcolor{black}{\begin{equation}\label{eq:P}
P=-\frac{M^2}{2r^4},
\end{equation}}
where
$M$ is a constant of integration related to the electric charge, and $\mathcal{H}(P)$ from (\ref{eq:H}) is a real function.
% represents the electric charge.
The electric field is obtained from the constitutive relations (\ref{eq:constitutive}),
$E \equiv F_{tr}= \mathcal{H}_p D$. Using expression (\ref{eq:H})
for $\mathcal{H}(P)$, the electromagnetic field strength results in
\begin{align}\label{eq:FmnE}
F_{\mu\nu}&=2\delta^t_{[\mu}\delta^r_{\nu]}E(r)&&\nonumber\\
&= 2\delta^{t}_{[\mu}\delta^{r}_{\nu]} \left(\frac {r\alpha_1}{l\kappa\sqrt {M}}-\frac{
\alpha_2}{\kappa}-\frac{l^2 M \left(
3\,\alpha_1\,\alpha_3-\alpha_2^2 \right) }{3 \kappa\,r^2}\right).&&
\end{align}
Notice that in order to recover the AdS spacetime asymptotically, the cosmological constant must take the following value,
\begin{equation}\label{eq:Lambda}
\Lambda=-\frac{3}{l^2}.
\end{equation}
Let us bring our attention to the fact that the difference between the temporal \textcolor{black}{$E_{t}^{t}$} and radial diagonal  \textcolor{black}{$E_r^r$} components of the
mixed version of Einstein's equations (\ref{eq:Einstein}) is proportional to the following fourth-order Cauchy-Euler ordinary differential equation
\begin{eqnarray}\label{eq:poly}
r^4 f^{(4)}+12r^3 f'''+36r^2f''+24rf'&=&0.
\end{eqnarray}
%The general solution of this equation (\ref{eq:poly}) is a superposition of negative powers of $r$, which are roots of the following characteristic polynomial
%\begin{eqnarray}
%m(m-1)(m-2)(m-3)=0.
%\end{eqnarray}
%obtained from substituting $h \varpropto 1/r^m$ in the differential equation (\ref{eq:poly}).
Therefore, the solution of the gravitational potential  is
\begin{eqnarray}\label{eq:f1}
f \left( r \right) =1-C_1{\frac {{l}}{r}}+ C_2{\frac {{l^2}}{{r}^{2}}}- C_3{\frac {{l^3}}{{r}^{3}}},
\end{eqnarray}
where the fourth integration constant is fixed to comply with the asymptotic behaviour ${\lim_{r\rightarrow +\infty}f(r)=1}$.
%$\lim_{r\rightarrow \infty}h(r)=0$, (which is equivalent to demanding $\lim_{r\rightarrow \infty}f(r)=1$).
%Next, we analyze $E^{\mu}_{\, \nu}$.
\textcolor{black}{Additionally, if we replace the expressions in (\ref{eq:f1}) and (\ref{eq:P}) in the equations of motion (\ref{eq:Einstein}) we obtain an equation to determine $\mathcal{H}_P$, which can be later integrated to obtain $\mathcal{H}(P)$, given previously in (\ref{eq:H})}. Finally, the remaining equations of motion are satisfied if the previous integration constants $C_i$'s from (\ref{eq:f1}) are fixed in terms of the charge-like parameter $M$ through the structural coupling constants as follows
\begin{eqnarray}\label{eq:Ps}
C_1=\alpha_1 \sqrt{M}\,,\,\, C_2= \alpha_2 M \,,\,\,\mathrm{and}\,\, C_3=\alpha_3 M^{3/2}\,,
\end{eqnarray}
{where the $\alpha_i$'s are in the appropriate units in order to the integration constants $C_i$'s be dimensionless.} The addition of the NLE \textcolor{black}{yields} to a rich structure for the metric function $f$ obtained previously in (\ref{eq:f1})-(\ref{eq:Ps}), where the uncharged case is recovered when $\alpha_1=\alpha_2=0$. This also shows that the linear Maxwell field scenario (that is $\mathcal{H}(P)=P$) is not allowed, which reinforces the necessity to explore other charged theories such as NLE. Moreover, from a physical perspective, this new structure for the metric function $f$ constructed via CG and NLE (\ref{eq:Squad}), will allow us to explore solutions with different numbers of horizons, in addition to nonzero thermodynamic quantities, as we will see bellow. These solutions are, to our knowledge, the first example of solutions in four-dimensional CG where their thermodynamic parameters do not vanish.
%Notice that the structure function does not satisfy the correspondence principle,
%that is, the non-linear electrodynamics considered does not approach Maxwell's
%theory at the weak field limit ($|l^2P| << 1$). This should not worry us as the proposed
%electrodynamics has no correspondence to Maxwell theory for weak fields.

%%%%%%%%%%%%%%%%%%%%%%%%%%%%%%%%%%%%%%%%%%%%%%%%%%%%%%%%%%%%%%%%%%%%%%%%%%%%%%%%%%%%%%%%%%%%
\section{Analysis of the solutions}
%%%%%%%%%%%%%%%%%%%%%%%%%%%%%%%%%%%%%%%%%%%%%%%%%%%%%%%%%%%%%%%%%%%%%%%%%%%%%%%%%%%%%%%%%%%%

We have established that the introduction of non-linear electrodynamics when considering CG, results in AdS solutions of the form (\ref{eq:ansatz})
where
\begin{eqnarray}\label{eq:f}
f \left( r \right) =1-\alpha_1\sqrt{M}{\frac {{l}}{r}}+\alpha_2 M {\frac {{l^2}}{{r}^{2}}}-\alpha_3 M^{3/2} {\frac {{l^3}}{{r}^{3}}},
\end{eqnarray}
provided that $\mathcal{H}(P)$ and $\Lambda$ are given by (\ref{eq:H}) and (\ref{eq:Lambda}) respectively. However, this set of expressions only represents a black hole solution if a horizon can be formed, that is, if there exists $r_h>0$ such that $f(r_h)=0$.
% That is, it is not enough to have
%\begin{eqnarray}
%r=
%\end{eqnarray}

To this effect, in the following subsections, we study the conditions in which eq. (\ref{eq:f}) can vanish, through the analysis of its asymptotic behavior as well as its maxima and minima.

First, let us notice that when { $r\rightarrow +\infty$}, $f(r)$ will approach 1 asymptotically and, in this regime, {$f(r)\simeq 1-\alpha_1 \sqrt{M} \frac{l}{r}$.} As a result, the sign of $\alpha_1$ will determine whether the function $f(r)$ approaches the horizontal asymptote from above or from below. Next, let us observe that when $r\rightarrow 0^+$, {$f(r)\simeq -  \alpha_3 M^{3/2}\frac{l^3}{r^3}$}, that is, the sign of $\alpha_3$ will determine if the function $f(r)$ starts increasing from $-\infty$ or decreasing from \textcolor{black}{$+\infty$} in the region {$r\in(0,+\infty)$.}

\textcolor{black}{Moreover, we can perform an analysis of the extreme values of $f(r)$. From the calculation of $f'(r)$ we find that $f(r)$ admits two extreme values which are located at
\begin{eqnarray}
r_{ext\,i}&=&{\frac { \sqrt{M}\,\left( { \alpha_2}\pm\sqrt {{{ \alpha_2}}^{2} -3\,{ \alpha_1}\,{
\alpha_3}}\right) l}{{ \alpha_1}}
},\label{eq:rext}
\end{eqnarray}
where the nature of these extreme values (whether they are a maximum or a minimum) will depend on the sign of their evaluation in $f''(r)$, that is
\begin{eqnarray}
f''(r_{ext\,i})=\pm \frac{
2\alpha_1^4\sqrt{\alpha_2^2-3\alpha_1 \alpha_3}}{(\sqrt{\alpha_2^2-3\alpha_1\alpha_3}\pm\alpha_2)^4 Ml^2}, \label{eq:Frr_e}
\end{eqnarray}
with $i=\{1,2\}$. Here, $r_{ext1}$ (resp.  $r_{ext2}$) is associated with positive (resp. negative) sign in equations (\ref{eq:rext}) and (\ref{eq:Frr_e}).
}
Let us notice that from \textcolor{black}{eqns. (\ref{eq:rext})-(\ref{eq:Frr_e}), } %(\ref{eq:Frr e1})-(\ref{eq:Frr e2}),
the existence of \textcolor{black}{real} extreme values is limited to the fulfillment of the condition
\begin{equation}\label{eq:condition}
\alpha_2^2-3\alpha_1\alpha_3 > 0.
\end{equation}
At this point, we are ready to analyze each case independently.

%%%%%%%%%%%%%%%%%%%%%%%%%%%%%%%%%%%%%%%%%%%%%%%%%
\subsection{Black holes with one horizon: The case $\alpha_1<0$ and $\alpha_3>0$}
%%%%%%%%%%%%%%%%%%%%%%%%%%%%%%%%%%%%%%%%%%%%%%%%%
{As previously mentioned, one can start by analyzing the asymptotical behavior of the function $f(r)$ in eq. (\ref{eq:f}). If one considers the limit ${r\rightarrow 0^+}$, the dominating term of $f(r)$ is $-\alpha_3 M^{3/2}\frac{l^3}{r^3}$. Assuming $M>0$ and $r>0$, then it is clear to see that for $\alpha_3>0$,  $\lim_{r \rightarrow 0^+} f(r) = -\infty$. On the other hand, if we analyze the regime $r \rightarrow +\infty$, we note that, in this limit, the dominant term of $f(r)$ is $1-\alpha_1 \sqrt{M} l/r$ which shows that, as $r$ increases, the function $f(r)$ will asymptotically approach a horizontal asymptote $f(r)=1$. In short, for the case $\alpha_3>0$, considering only positive values for $r$, the function $f(r)$ starts in the fourth quadrant and it increases for small values of $r$, since the function $f(r)$ asymptotically approaches the value of $1$ as $r$ approaches infinity, it must cross the horizontal axis, ensuring the existence of an event horizon $r_h>0$.
Moreover, if we consider $\alpha_1<0$, the function $f(r)$ will asymptotically approach the horizontal asymptote $f(r)=1$ from above. Also, after analyzing the first and second derivatives of $f(r)$, we note that the choice for the sign of $\alpha_1<0$ and $\alpha_3>0$ will result in the function $f(r)$ displaying an absolute maximum in the regime $r>0$, and the existence of a single horizon, regardless of the sign of $\alpha_2$, as seen in Fig. \ref{fig:onehorizon}. { This analysis is deeply studied in the Appendix \ref{appendix1}.}
}
\begin{figure}[h!]
\centering
\includegraphics[width=7cm]{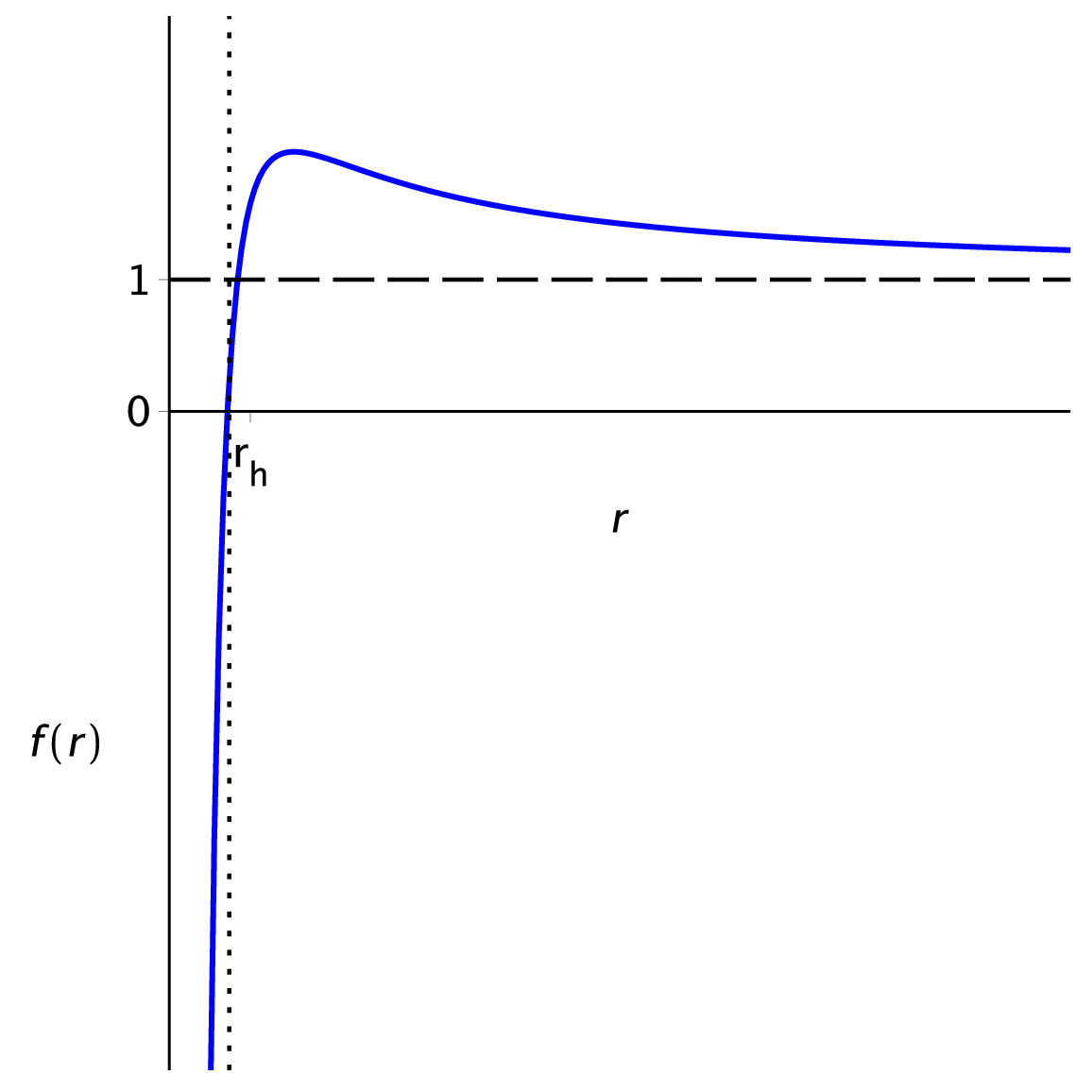}
\caption{Gravitational potential $f(r)$ associated to black holes with a single horizon when  $\alpha_1<0,\alpha_3>0$.}\label{fig:onehorizon}
\end{figure}

%\subsection{$\alpha_1>0,\alpha_3>0$}
%\begin{figure}[h!]
%  \centering
%    \includegraphics[scale=.37]{a1p_a3p.eps}
%      \caption{Generic example with $\alpha_1>0,\alpha_3>0$. }
%  \label{fig:a1p_a3p}
%\end{figure}

%\subsection{$\alpha_1<0,\alpha_3>0$}
%The black hole in figure \ref{fig:a1n_a3p} has one horizon.
%\begin{figure}[h!]
%  \centering
%    \includegraphics[scale=.37]{a1n_a3p.eps}
%      \caption{Generic example with $\alpha_1<0,\alpha_3>0$.}
%  \label{fig:a1n_a3p}
%\end{figure}
%%%%%%%%%%%%%%%%%%%%%%%%%%%%%%%%%%%%%%%%%%%%%%%%%
\subsection{Black holes with up to two horizons: The case $\alpha_3<0$}
%%%%%%%%%%%%%%%%%%%%%%%%%%%%%%%%%%%%%%%%%%%%%%%%%

On the contrary, when $\alpha_3<0$, the gravitational potential $f(r)$ is initially decreasing in the region $r>0$, one can face three scenarios: having two horizons, the  {extremal} case of one horizon, or no horizon at all, which does not represent a black hole.
%%%%%%%%%%%%%%%%%%%%%%%%%%%%%%%%%%%%%%%%%%%%%%%%%%%%%%%%%%%%%%%%%%%%%%%%%%%%%%%%%%%%%%%%%%%%%%%%%%%%%%%%%%%%%%%%%%
Landing on one case or another depends on the relations between $\alpha_1$, $\alpha_2$ and $\alpha_3$. Eq.(\ref{eq:f}) will represent the gravitational potential of a black hole only provided that
\begin{align}\label{eq:criticalbh}
4\alpha_1^3 \alpha_3-\alpha_1^2 \alpha_2^2-18 \alpha_1 \alpha_2 \alpha_3+4 \alpha_2^3+27 \alpha_3^2 \leq 0.
\end{align}
When the strict inequality is met, the solution will have two horizons (where the minimum $f(r_{min})<0$ is situated at $r_{min}>0$). Otherwise, when the equality is met, the solution will correspond to {an extremal configuration}, in which the minimum of $f(r_{e})=0$ is located at $r_e>0$. Moreover, according to \textcolor{black}{eqn. (\ref{eq:Frr_e})}, when $\alpha_1>0$, the function $f(r)$ will showcase a minimum; on the other hand, when $\alpha_1<0$, the function $f(r)$ will have both a minimum and a maximum, as seen in Fig. \ref{fig:twohorizons}. {This study is analyzed in Appendix \ref{appendix2}}.

\begin{figure}[h!]
\centering
\begin{minipage}[c]{7cm}
\includegraphics[width=7cm]{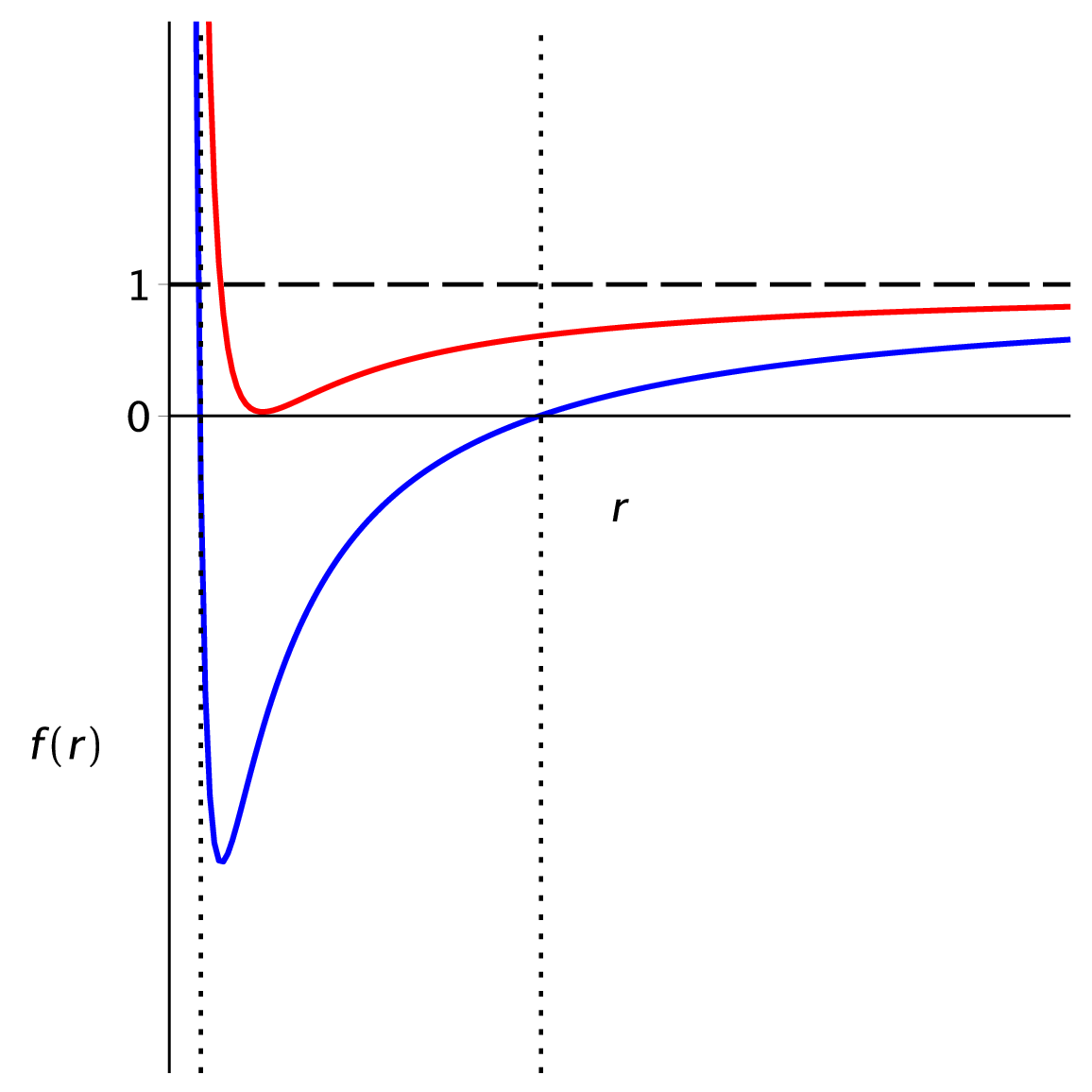}
\end{minipage}
\begin{minipage}[r]{7cm}
\includegraphics[width=7cm]{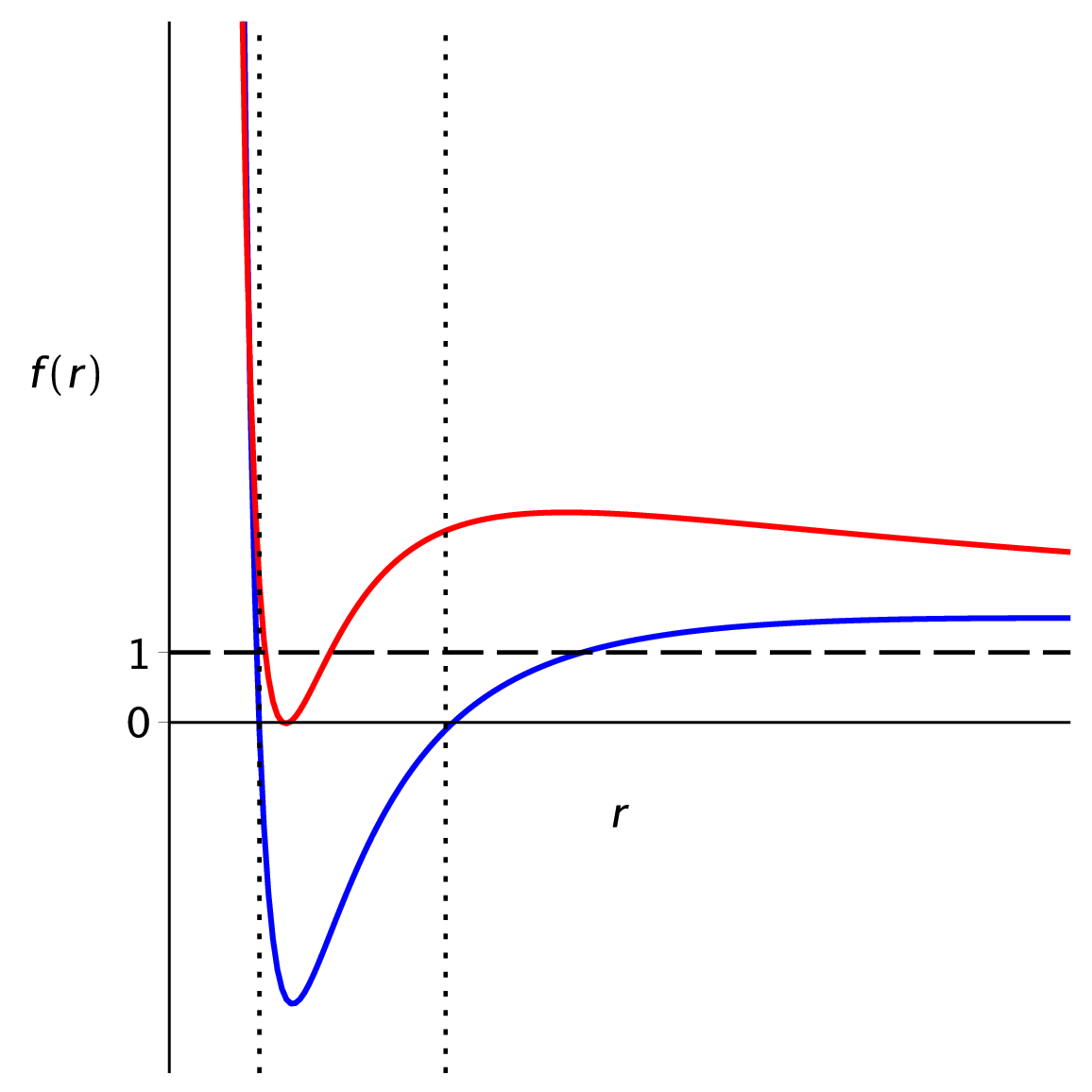}
\end{minipage}
\caption {{The graphs represent the gravitational potential $f(r)$ when having a minimum. The black graphs represent two-horizons black holes and the red graphs represent the extremal configurations}. Top: $f(r)$ when $\alpha_1>0,\alpha_3<0$. Bottom: $f(r)$ when  $\alpha_1<0,\alpha_3<0$.}\label{fig:twohorizons}
\end{figure}

%\subsection{$\alpha_1>0,\alpha_3<0$}
%As represented in figure \ref{fig:a1p_a3n}, the case $\alpha_1>0,\alpha_3<0$ may result in black holes with two horizons or, the critical case with one horizon (or no horizon at all for certain combinations of $\alpha_i$).The solution has two horizons when $4\alpha_1^3 \alpha_3-\alpha_1^2 \alpha_2^2-18 \alpha_1 \alpha_2 \alpha_3+4 \alpha_2^3+27 \alpha_3^2<0$ and the critical case arises when $4\alpha_1^3 \alpha_3-\alpha_1^2 \alpha_2^2-18 \alpha_1 \alpha_2 \alpha_3+4 \alpha_2^3+27 \alpha_3^2=0$. \textbf{CHECK}

%\begin{figure}[h!]
%  \centering
%    \includegraphics[scale=.37]{a1p_a3n.eps}
%      \caption{Black hole solutions with $\alpha_1>0,\alpha_3<0$.}
%  \label{fig:a1p_a3n}
%\end{figure}

%\subsection{$\alpha_1<0,\alpha_3<0$}
%As represented in figure \ref{fig:a1n_a3n}, the case $\alpha_1<0,\alpha_3<0$ may result in black holes with two horizons or, the critical case with one horizon (or no horizon at all for certain combinations of $\alpha_i$). The solution has two horizons when $4\alpha_1^3 \alpha_3-\alpha_1^2 \alpha_2^2-18 \alpha_1 \alpha_2 \alpha_3+4 \alpha_2^3+27 \alpha_3^2<0$ and the critical case arises when $4\alpha_1^3 \alpha_3-\alpha_1^2 \alpha_2^2-18 \alpha_1 \alpha_2 \alpha_3+4 \alpha_2^3+27 \alpha_3^2=0$. \textbf{CHECK}

%\begin{figure}[h!]
%  \centering
%    \includegraphics[scale=.37]{a1n_a3n.eps}
%      \caption{Black hole solutions with $\alpha_1<0,\alpha_3<0$.}
%  \label{fig:a1n_a3n}
%\end{figure}

%%%%%%%%%%%%%%%%%%%%%%%%%%%%%%%%%%%%%%%%%%%%%%%%%
\subsection{Black holes with up to three horizons: The case $\alpha_1>0$ and $\alpha_3>0$}
\begin{figure}[h!]
\centering
\begin{minipage}[c]{7cm}
\includegraphics[width=6cm]{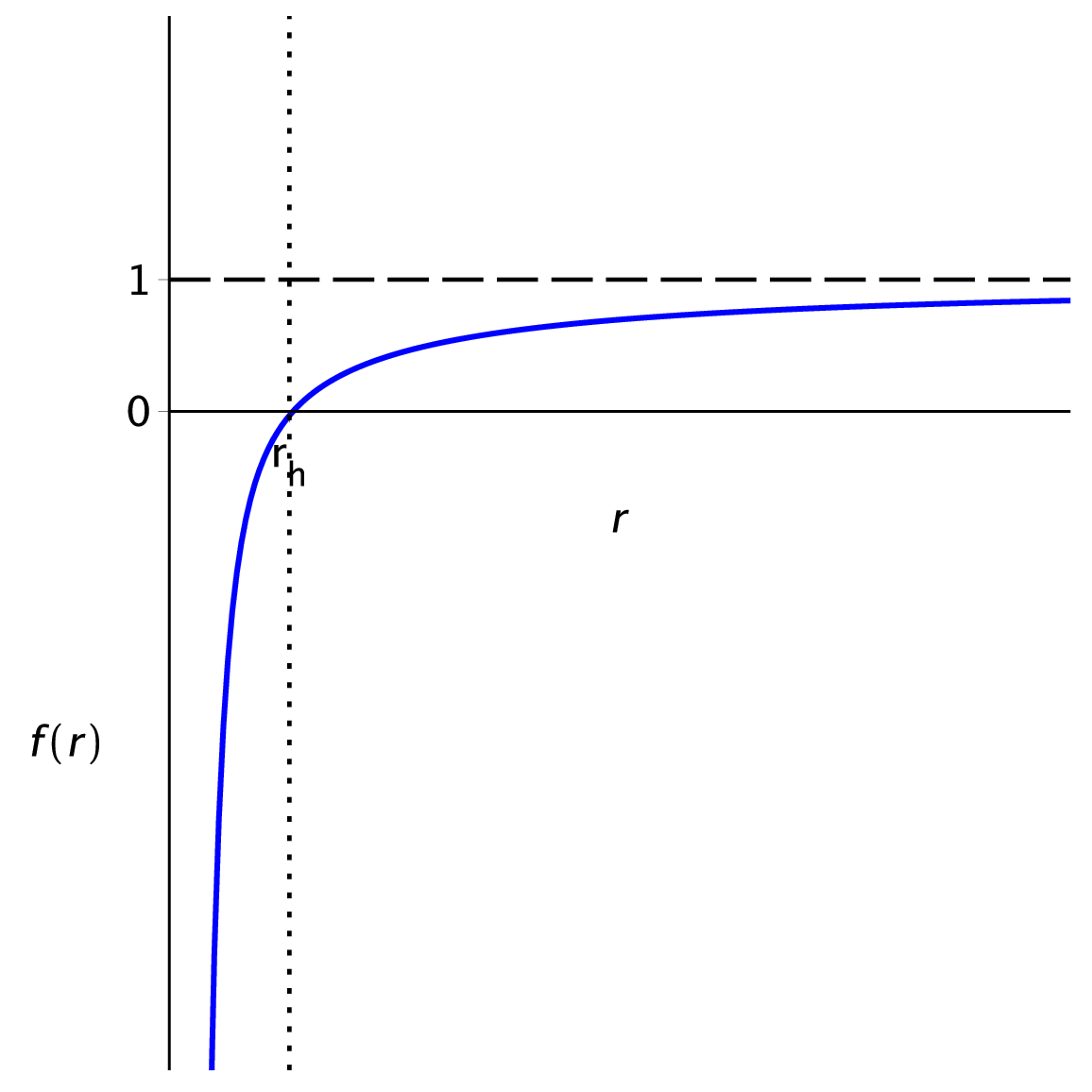}
\end{minipage}
\begin{minipage}[r]{7cm}
\includegraphics[width=7cm]{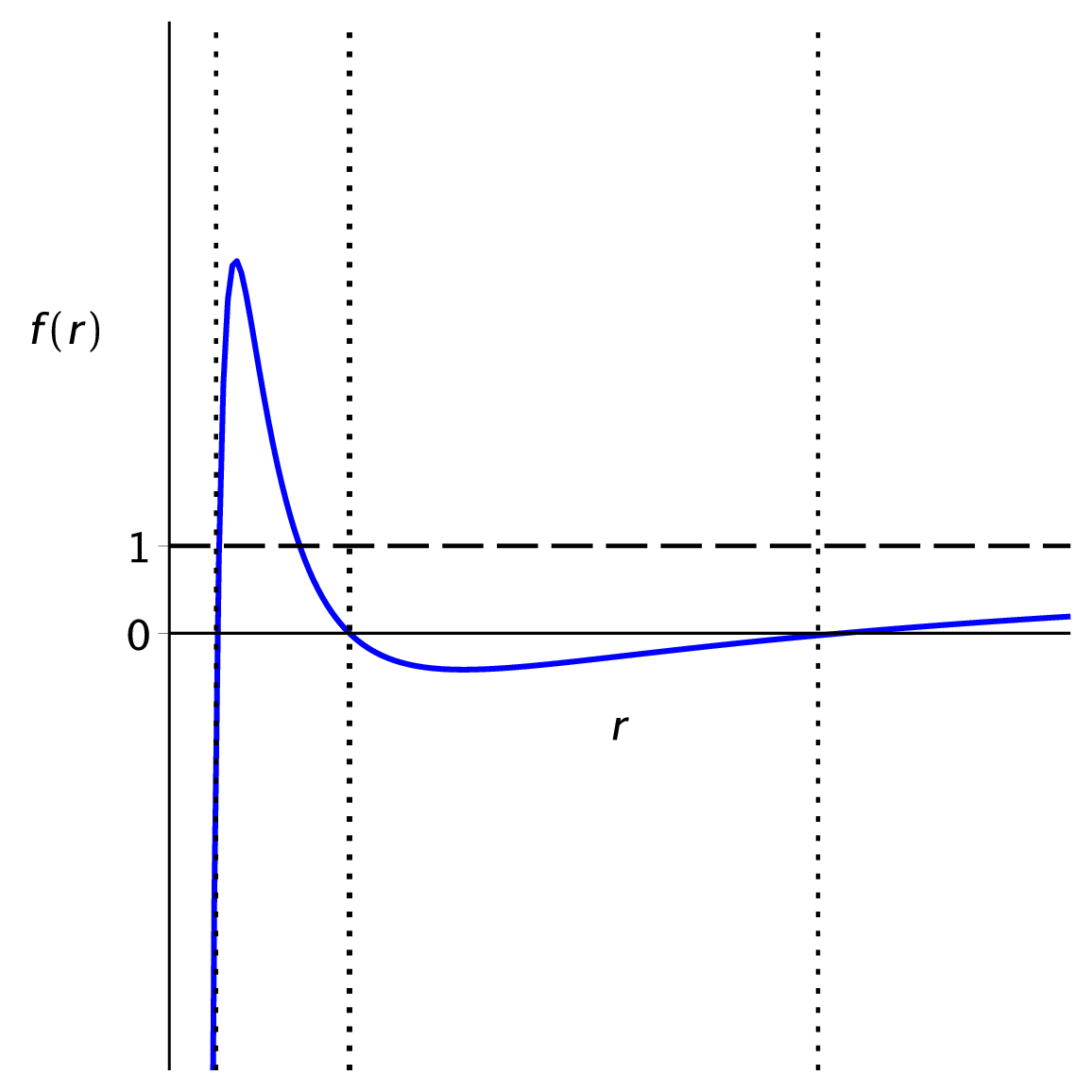}
\end{minipage}
\caption{Gravitational potential $f(r)$ associated to black holes when $\alpha_1>0,\alpha_3>0$. Top: solution with a single horizon when the condition $\alpha_2>0,\, \alpha_2^2-3\alpha_1 \alpha_3>0$ is not met. Bottom: solution with three horizons when the condition $\alpha_2>0,\, \alpha_2^2-3\alpha_1 \alpha_3>0$ is met.}\label{fig:threehorizons}
\end{figure}

%%%%%%%%%%%%%%%%%%%%%%%%%%%%%%%%%%%%%%%%%%%%%%%%%
Finally, when we consider the case $\alpha_1>0$ and $\alpha_3>0$. For very small but positive values of $r$, we notice that the function $f(r)$ is increasing from $-\infty$ while, for large values of $r$ (that is $r\rightarrow +\infty$), the function $f(r)$ approaches the value of 1 from below. This ensures the existence of at least one horizon (in the region $r>0$). However, under certain circumstances we observe that this case can admit {up to} three horizons, exhibiting one maximum and one minimum in the region $r>0$. The conditions that will determine which case we will land in are detailed in the Appendix, but let us now state, in advance, that when the conditions $\alpha_1>0$, $\alpha_2>0$, $\alpha_3>0$ and $\alpha_2^2-3\alpha_1 \alpha_3>0$ are met, we will have a solution with three horizons. In Fig. \ref{fig:threehorizons}, we show examples of both cases for clarity. However, it is interesting to remark that, this case also admits two horizons when the maximum corresponds to the internal horizon or when the minimum corresponds with the outer horizon, as seen in Fig. \ref{fig:threehorizons_critical}. This scenario is fully explored in the Appendix \ref{app:3horizons}.

%%%%%%%%%%%%%%%%%%%%%%%%%%%%%%%%%%%%%%%%%%%%%%%%%%%%%%%%%%%%%%%%%%%
\section{Thermodynamics and local stability of the solution }
%%%%%%%%%%%%%%%%%%%%%%%%%%%%%%%%%%%%%%%%%%%%%%%%%%%%%%%%%%%%%%%%%%%

Given the structure of these new nonlinearly charged black hole solutions in four dimensions, it is interesting to explore their thermodynamics. As a first step, we will consider the electric charge which reads
\textcolor{black}{\begin{eqnarray}\label{eq:charge-4d}
\mathcal{Q}_{e}=\int d \Omega_{2} \left(\frac{r}{l}\right)^{2}n^{\mu} u^{\nu} P_{\mu \nu}=\frac{M \Omega_{2}}{l^2}=\frac{\Omega_{2} r_h^{2}}{\zeta^{2} l^{4}},
\end{eqnarray}}
where $r_h$ is the location of the event (or outer) horizon
%depending on the election for the coupling constants $\alpha_1$, $\alpha_2$ and $\alpha_3$
which can be expressed as $r_h=\zeta \sqrt{M}l$, where $\zeta$ is a root of the \textcolor{black}{cubic polynomial}
\begin{equation}\label{eq:pol}
\zeta^3-\alpha_1\zeta^2+\alpha_2\zeta-\alpha_3=0;
\end{equation}
\textcolor{black}{$\Omega_{2}$ is the finite volume of the compact planar base manifold given by $\int d x_1 d x_2= \int d \Omega_{2}=\Omega_{2}=\Omega_{x_1} \Omega_{x_2}$, } while $n^{\mu}$  and $u^{\nu}$ are the unit spacelike and timelike normals to a sphere of radius $r$ given by
\textcolor{black}{\begin{eqnarray}\label{eq:vectors}
{n^{\mu}:=\frac{dt}{\sqrt{-g_{tt}}}=\frac{l}{r \sqrt{f}}dt,\quad u^{\mu}:=\frac{dr}{\sqrt{g_{rr}}}=\frac{r \sqrt{f}}{l}dr.}\quad
\end{eqnarray}}
As a first step to calculate the electric potential we must determine the \textcolor{black}{4-potential} $A_{\mu}$. We achieve this by integrating eq. (\ref{eq:FmnE}) and taking into account that $F_{\mu\nu}=2\partial_{[\mu}A_{\nu]}$. As a result the only non-zero component of the \textcolor{black}{4-potential} is given by

%Together with the above, from the expressions of $P_{\mu \nu}$ and $\mathcal{H}(P)$ given previously in (\ref{eq:Pmunu}) and (\ref{eq:H}), we have
{\begin{eqnarray}\label{vect-pot-d4}
A_{t}(r)={\frac {{r}^{2}{\alpha_1}}{2 \sqrt {M}l\kappa}}-{\frac {
{\alpha_2}\,r}{\kappa}}+{\frac {{l}^{2}M \left(
3\,{\alpha_1}\,{\alpha_3}-{{\alpha_2}}^{2} \right) }{3 \kappa\,r}},\quad
\end{eqnarray}
where the integration constant in our case is null,} and the electric potential \footnote{In this work we have used the same definition of the electric potential as \cite{Liu:2014dva,Bravo-Gaete:2020ftn,Bravo-Gaete:2015xea}.} is given by:
\begin{eqnarray}\label{pot-4d}
\Phi_{e}=-A_{t}(r)\Big{|}_{r=r_h}&=&-{\frac {3{\alpha_1}\,r_h^{2}}{2\sqrt {M}l\kappa}}+{\frac
{ \left(\alpha_1^2+\alpha_2 \right) r_h}{\kappa}}\nonumber \\
&-&{
\frac {{\alpha_1}\,{\alpha_2}\,\sqrt {M}l}{\kappa}}+
{\frac {{l}^{2}M \alpha_2^{2}}{3 \kappa\, r_h}},\nonumber \\
&=&\frac{r_h}{\kappa} \Big({\alpha_2}+
\alpha_1^{2}-\frac{3}{2} \alpha_1\zeta\nonumber\\
&-&
{\frac {\alpha_1\,\alpha_2}{\zeta}}+\frac{1}{3}\,
{\frac {\alpha_2^{2}}
{\zeta^{2}}}
\Big).
\end{eqnarray}
On the other hand, to compute the entropy we will consider Wald's formula \cite{Wald:1993nt,Iyer:1994ys} which, in our case, yields to

\textcolor{black}{\begin{eqnarray}
\label{wald} \mathcal{S}_{W}&{:=}&-2 \pi \int_{{H}} d^{2}x \sqrt{|h|} \left(\frac{\delta \mathcal{L}_{\mathrm{grav}}}{\delta R_{\mu \nu \sigma \rho}} \, \varepsilon_{\mu \nu} \,  \varepsilon_{\sigma \rho}\right),\nonumber\\
 &=&\frac {2  \left( 3\,\alpha_{1} \sqrt{M} r_h-2\,
\alpha_{2} Ml \right) \pi \Omega_{2} }{3\,\kappa\,l},\nonumber\\
&=&\frac{2 \Omega_{2} \pi}{\kappa} \left(\frac{r_h}{l}\right)^{2}\left(\frac{\alpha_1}{\zeta}-\frac{2 \alpha_2}{3 \zeta^{2}}\right),
\end{eqnarray}
where
\begin{eqnarray*}
\frac{\delta \mathcal{L}_{\mathrm{grav}}}{\delta
R_{\alpha \beta \gamma \delta}}&=&\frac{1}{4 \kappa}\,\Big(g^{\alpha\gamma}g^{\beta\delta}-g^{\alpha\delta}g^{\beta\gamma}\Big)
\nonumber\\
&+&\frac{\beta_1}{2 \kappa}\, R \left(g^{\alpha \gamma } g^{\beta
\delta } -g^{\alpha \delta } g^{\beta \gamma
}\right)\nonumber\\
&+&\frac{\beta_{2}}{4 \kappa}\, \left(g^{\beta \delta } R^{\alpha
\gamma }-g^{\beta \gamma } R^{\alpha \delta } -g^{\alpha \delta }
R^{\beta \gamma }+g^{\alpha \gamma } R^{\beta \delta
}\right),
\end{eqnarray*}
with $\beta_1$ and $\beta_2$ subject to (\ref{relations}), and the integral is evaluated on a 2-dimensional spacelike surface $H$ (the bifurcation surface)  characterized by the fact that the timelike Killing vector {$\partial_{t}=\xi^{\mu}\partial_{\mu}$} vanishes, $|h|$ denotes the determinant of the induced metric on ${H}$, $\varepsilon_{\mu \nu}$ represents the binormal antisymmetric tensor constructed via the wedge product of the unit normal vectors $n^{\mu}$ and $u^{\mu}$ from (\ref{eq:vectors}), normalized as $\varepsilon_{\mu \nu} \varepsilon^{\mu \nu}=-2$.}
Additionally, the Hawking temperature takes the form
\begin{eqnarray}\label{eq:Temp}
T{:=}\frac{k}{2 \pi}\Big{|}_{r=r_h}&=&\frac{3 r_h}{4 \pi l^{2}}-\frac{\alpha_1 \sqrt{M}}{2 \pi l}+\frac{\alpha_2 M}{4 \pi r_h},\nonumber\\
&=&\frac{r_h}{4 \pi l^{2}} \left(3-\frac{2 \alpha_1}{\zeta}+\frac{\alpha_2}{\zeta^{2}}\right),
\end{eqnarray}
\textcolor{black}{where $k$ is the surface gravity which reads
$$k=\sqrt{-\frac{1}{2}\left(\nabla_{\mu} \xi_{\nu}\right)\left(\nabla^{\mu} \xi^{\nu}\right)}.$$}
Finally, to calculate the mass of these charged AdS black hole configurations we will consider the approach described in \cite{Kim:2013zha,Gim:2014nba}, corresponding to an off-shell prescription of the Abbott-Desser-Tekin (ADT) procedure \cite{Abbott:1981ff,Deser:2002rt,Deser:2002jk}. The choice of this method to calculate conserved charges is ideal for CG due to the presence of quadratic curvature terms in its gravitational action.

%and given by the presence of quadratic corrections in the gravitational action (\ref{eq:Squad}) with the relations (\ref{relations}), we will consider the approach described in \cite{Kim:2013zha,Gim:2014nba}, corresponding to a off-shell prescription of the Abbott-Deser-Tekin (ADT) \cite{Abbott:1981ff,Deser:2002rt,Deser:2002jk} procedure.
The main elements of the quasilocal method are the surface term
\begin{eqnarray}
\Theta^{\mu}&=&2\sqrt{-g}\Biggl[\left(\frac{\delta \mathcal{L}_{\mathrm{grav}}}{\delta R_{\mu \alpha \beta \gamma}}\right)\nabla_{\gamma}\delta g_{\alpha\beta}-\delta g_{\alpha\beta} \nabla_{\gamma}\left(\frac{\delta \mathcal{L}_{\mathrm{grav}}}{\delta R_{\mu \alpha \beta \gamma}}\right)\nonumber\\
\nonumber\\
&+&\frac{1}{2}\,\left(\frac{\delta \mathcal{L}_{NLE}}{\delta \left(\partial_{\mu} A_{\nu}\right)}\right)\delta A_{\nu}\Biggr],
\end{eqnarray}
and the Noether potential
\begin{eqnarray}
K^{\mu\nu}&=&\sqrt{-g}\Bigg[2\left(\frac{\delta \mathcal{L}_{\mathrm{grav}}}{\delta R_{\mu \nu \rho \sigma}}\right)\nabla_{\rho}\xi_{\sigma} - 4\xi_{\sigma}\nabla_{\rho}\left(\frac{\delta \mathcal{L}_{\mathrm{grav}}}{\delta R_{\mu \nu \rho \sigma}}\right)\nonumber\\
&-&\left(\frac{\delta \mathcal{L}_{NLE}}{\delta \left(\partial_{\mu} A_{\nu}\right)}\right) \xi^{\sigma} A_{\sigma}\Bigg].
\end{eqnarray}
With all the above, using a parameter $s \in [0,1]$, we interpolate between the charged solution at $s=1$ and the asymptotic one at $s=0$, resulting in  the quasilocal charge:
\begin{eqnarray*}
{\mathcal{M}(\xi)=\int_B d^{2}x_{\mu\nu}\left(\Delta K^{\mu \nu}(\xi)-2\xi^{[\mu} \int^{1}_0 ds \Theta^{\nu]}\right),}
\end{eqnarray*}
where $\Delta K^{\mu\nu}(\xi)\equiv K^{\mu\nu}_{s=1}(\xi)-K^{\mu\nu}_{s=0}(\xi)$ is the difference of the Noether potential between the interpolated solutions. For this particular case the mass reads
\begin{eqnarray}
\label{mass-d4}
\mathcal{M}&=&\frac{\alpha_1 \alpha_2 M^{\frac{3}{2}}\Omega_{2}}{9 l \kappa}=
\frac{\alpha_1 \alpha_2 r_h^{3}\Omega_{2}}{9 l^{4} \kappa \zeta ^{3}}.
\end{eqnarray}
Notice that for the Wald entropy $\mathcal{S}_{W}$ (\ref{wald}), as well as the mass  $\mathcal{M}$ (\ref{mass-d4}), the presence of the coupling constants $\alpha_1$ and $\alpha_2$ is providential, reinforcing the importance of the non-linear electrodynamic as a matter source with this gravity theory. In fact, when $\alpha_1=\alpha_2=0$, the vector potential $A_{t}(r)$ (\ref{vect-pot-d4}) vanishes, recovering the well known four-dimensional Schwarzschild- AdS black hole with a planar base manifold in CG, whose extensive thermodynamical quantities $\mathcal{M}$ and $\mathcal{S}_{W}$ are null, while  the Hawking Temperature is $T=3r_h /(4 \pi l^2)$.

\textcolor{black}{Just for completeness, from eqns. (\ref{mass-d4}), (\ref{wald}) and (\ref{eq:charge-4d}) we have:
\begin{eqnarray*}
\delta \mathcal{M}&=&
\frac{\alpha_1 \alpha_2 r_h^{2}\Omega_{2}}{3 l^{4} \kappa \zeta ^{3}}\,\delta r_h,\\
\delta \mathcal{S}_{W}&=&\frac{4 \Omega_{2} \pi r_h}{\kappa l^2} \left(\frac{\alpha_1}{\zeta}-\frac{2 \alpha_2}{3 \zeta^{2}}\right)\,\delta r_h,\\
\delta \mathcal{Q}_{e}&=&\frac{2 \Omega_{2} r_h}{\zeta^{2} l^{4}} \,\delta r_h,
\end{eqnarray*}
and together with the electric potencial $\Phi_{e}$ (\ref{pot-4d}) as well as the Hawking temperature $T$ (\ref{eq:Temp}), a first law of the black holes thermodynamics
\begin{eqnarray}\label{eq:first-law}
\delta \mathcal{M}= T \delta \mathcal{S}_{W}+\Phi_e  \delta \mathcal{Q}_{e},
\end{eqnarray}
arises. Together with the above, through the thermodynamical parameters (\ref{eq:charge-4d}), (\ref{pot-4d})-(\ref{eq:Temp}), we can express the mass (\ref{mass-d4}) as a function of the extensive thermodynamical quantities $\mathcal{S}_{W}$ and $\mathcal{Q}_{e}$ in the following form
\begin{eqnarray}
\mathcal{M}(\mathcal{S}_{W},\mathcal{Q}_{e})&=&{\frac {\sqrt {6}\,\sqrt {\kappa}\,{\mathcal{S}_{W}}^{3/2} \left( 3\,{\zeta }^{2}-
2\,\alpha_{1} \zeta +\alpha_{2} \right) }{ 12 \sqrt {
\Omega_{{2}}}{\pi }^{3/2}\,\sqrt {3\,\alpha_{1}\zeta
-2\,\alpha_{2}}\,l \zeta }}\nonumber \\
&+&\frac{\mathcal{Q}_{e}^{3/2} l^{2} \zeta \Psi}{9 \sqrt{\Omega_{2}} \kappa},\label{eq:mass(s,q)}
\end{eqnarray}
with
\begin{equation}
\Psi=6
\,\alpha_{2}+6\alpha_{1}^{2}-9\,\alpha_{1}\zeta-{\frac {6 \alpha_{2}
\alpha_{1}}{\zeta}}+\frac {2 \alpha_{2}^{2}}{{\zeta}^{2}},\label{eq:Psi}
\end{equation}
where it is straightforward to verify that the intensive parameters
$$T=\left(\frac{\partial \mathcal{M}}{ \partial \mathcal{S}_{W}}\right)_{{\mathcal{Q}}_{e}}, \qquad \Phi_{e}=\left(\frac{\partial \mathcal{M}}{ \partial \mathcal{Q}_{e}}\right)_{{\mathcal{S}}_{W}},$$
are consistent with the {expressions}  (\ref{pot-4d})  and (\ref{eq:Temp}) (where the subindices stand for \emph{at constant electric charge} ${{\mathcal{Q}}_{e}}$, and \emph{at constant entropy} ${{\mathcal{S}}_{W}}$ respectively). Additionally, under a rescaling with a nonzero parameter $\lambda$, equation (\ref{eq:mass(s,q)}) becomes
$$\mathcal{M}(\lambda \mathcal{S}_{W}, \lambda \mathcal{Q}_{e})= \lambda^{\frac{3}{2}} \mathcal{M}(\mathcal{S}_{W},\mathcal{Q}_{e}), $$
yielding to a four-dimensional Smarr formula \cite{Smarr:1972kt}
\begin{eqnarray}\label{eq:Smarr-4d}
\mathcal{M}=\frac{2}{3} \left(T \mathcal{S}_{W}+\Phi_e \mathcal{Q}_{e}\right),
\end{eqnarray}
which corresponds to a particular case of the higher-dimensional situation \cite{Dehghani:2013mba}
\begin{eqnarray}\label{eq:Smarr-d}
\mathcal{M}=\left(\frac{D-2}{D-1}\right) \left(T \mathcal{S}_{W}+\Phi_e \mathcal{Q}_{e}\right),
\end{eqnarray}
where $D$ is the dimension of the space-time, highly explored in \cite{Bravo-Gaete:2015iwa,Bravo-Gaete:2021hza,Liu:2015tqa}.
}

Given these thermodynamical quantities, it is interesting to study this system under small perturbations around the equilibrium. In our case, we will consider the {\em{grand canonical ensemble}}, where the intensive thermodynamical quantities are fixed. With this, we can express the entropy, mass, and charge in functions of $T$ and $\Phi_{e}$ in the following form
\begin{eqnarray}
\mathcal{S}_{W}&=&{\frac {32}{3}}\,{\frac {{l}^{2}{T}^{2}\Omega_{{2}}{\pi }^{3}{\zeta}^{
2} \left( 3\,\alpha_{{1}}\zeta-2\,\alpha_{{2}} \right) }{ \left( 3\,{
\zeta}^{2}-2\,\alpha_{{1}}\zeta+\alpha_{{2}} \right) ^{2}\kappa}},\\
\mathcal{Q}_{e}&=&\frac{36\,\Phi_{e}^{2}\kappa^{2}\Omega_{2}}
{\Psi^{2} \zeta^2 {l}^{4}},\\
\mathcal{M}&=&{\frac {64}{9}}\,{\frac {{l}^{2}{T}^{3}\Omega_{{2}}{\pi }^{3}{\zeta}^{
2} \left( 3\,\alpha_{{1}}\zeta-2\,\alpha_{{2}} \right) }{ \left( 3\,{
\zeta}^{2}-2\,\alpha_{{1}}\zeta+\alpha_{{2}} \right) ^{2}\kappa}}\nonumber\\
&+&\frac{24\,\Phi_{e}^{3}\kappa^{2}\Omega_{2}}
{\Psi^{2} \zeta^2 {l}^{4}}.
\end{eqnarray}
With this information, we are in a position to determine the local thermodynamical (in)stability of this charged black hole solution {under thermal fluctuations} through the behavior of the specific heat $C_{\Phi_{e}}$, given by
\begin{eqnarray}\label{eq:cq}
C_{\Phi_{e}}&=&\left(\frac{\partial \mathcal{M}}{\partial T}\right)_{\Phi_{e}}=T \left(\frac{\partial \mathcal{S}_{W}}{\partial T}\right)_{\Phi_{e}}\nonumber\\
&=&{\frac {64}{3}}\,{\frac {{l}^{2}{T}^{2}\Omega_{{2}}{\pi }^{3}{\zeta}^{
2} \left( 3\,\alpha_{{1}}\zeta-2\,\alpha_{{2}} \right) }{ \left( 3\,{
\zeta}^{2}-2\,\alpha_{{1}}\zeta+\alpha_{{2}} \right) ^{2}\kappa}}.
\end{eqnarray}
Here we observe that for $T\geq 0$ {or in the same way,
\begin{equation}
\Psi_1:=3 \zeta^2-2\alpha_1\zeta+\alpha_2 \geq 0, \label{cond1}
\end{equation}
the specific heat becomes {non-negative when
\begin{equation}\label{eq:cond}
\Psi_2:=3\,\alpha_{{1}}\zeta-2\,\alpha_{{2}} \geq 0,
\end{equation}
which} can be interpreted as a locally stable configuration. Nevertheless, it is worth pointing out that, to have a real {and well-defined mass} according to the expression (\ref{eq:mass(s,q)}) as well as specific heat (\ref{eq:cq}), only the strict inequalities from (\ref{cond1}) and  (\ref{eq:cond}) are considered .} Therefore, here we can conclude that the introduction of the nonlinear electrodynamics to {the} CG action induces rich thermodynamical properties. The comment on stability above is consistent with the analysis of the Gibbs Free Energy $G(T,\Phi_{e})=\mathcal{M}-T \mathcal{S}_{W}-\Phi_{e} \mathcal{Q}_{e}$, given by
\begin{eqnarray}\label{eq:GFE}
G(T,\Phi_{e})&=&-{\frac {32}{9}}\,{\frac {{l}^{2}{T}^{3}\Omega_{{2}}{\pi }^{3}{\zeta}^{
2} \left( 3\,\alpha_{{1}}\zeta-2\,\alpha_{{2}} \right) }{ \left( 3\,{
\zeta}^{2}-2\,\alpha_{{1}}\zeta+\alpha_{{2}} \right) ^{2}\kappa}}\\
&-&\frac{12\,\Phi_{e}^{3}\kappa^{2}\Omega_{2}}
{\Psi^{2} \zeta^2 {l}^{4}},\nonumber
\end{eqnarray}
where its Hessian matrix $H_{ab}:= \partial_{a} \partial_{b} G(T,\Phi_{e})$, { with $a,b \in \{T,\Phi_{e}\}$,}  satisfies the conditions
\begin{eqnarray*}
H_{TT}&=&-{\frac {64}{3}}\,{\frac {{l}^{2}{T}\Omega_{2}{\pi }^{3}{\zeta}^{
2} \left( 3\,\alpha_{{1}}\zeta-2\,\alpha_{{2}} \right) }{ \left( 3\,{
\zeta}^{2}-2\,\alpha_{{1}}\zeta+\alpha_{{2}} \right) ^{2}\kappa}} \leq 0,\\
H_{\Phi_{e} \Phi_{e}}&=&-\frac{72\,\Phi_{e} \kappa^{2}\Omega_{2}}
{\Psi^{2} \zeta^2
{l}^{4}}\leq 0,\\
|H_{a b}|&=&H_{TT} H_{\Phi_{e} \Phi_{e}}- \left(H_{T \Phi_{e}}\right)^{2}\\
&=& \frac{1536\, \Omega_{2}^{2}\,T \,\pi^{3} \left( 3\,\alpha_{1}
\zeta-2\,\alpha_{2} \right) \kappa\,\Phi_{e}}{{l}^{2} \left(3\,{\zeta}^{
2}-2\,\alpha_{1}\zeta+\alpha_{2} \right) ^{2} \Psi ^{2}}\geq 0,\\
\end{eqnarray*}
if $T,\Phi_{e} \geq 0$, {this is if (\ref{cond1})-(\ref{eq:cond}) and
\begin{eqnarray}
&&\Psi=\frac{\alpha_1 \alpha_2}{\zeta}-\frac{\Psi_1\Psi_2}{\zeta^2} \geq 0,  \label{cond2}
\end{eqnarray}
hold, where $\Psi$ was defined previously in (\ref{eq:Psi}) and in order to have a well-defined Gibbs free energy (\ref{eq:GFE}), we consider only the strict inequality from (\ref{cond2}). As an example, for $\zeta=1$, which implies that $r_h=\sqrt{M}l$, we have that the strict inequalities (\ref{cond1})-(\ref{eq:cond}) and (\ref{cond2}) are satisfied when the constants $\alpha_1$ and $\alpha_2$ belong to the region $\cal{R}$ represented in the Figure \ref{fig:figR}.
\begin{figure}[h!]
 \centering
    \includegraphics[scale=.08]{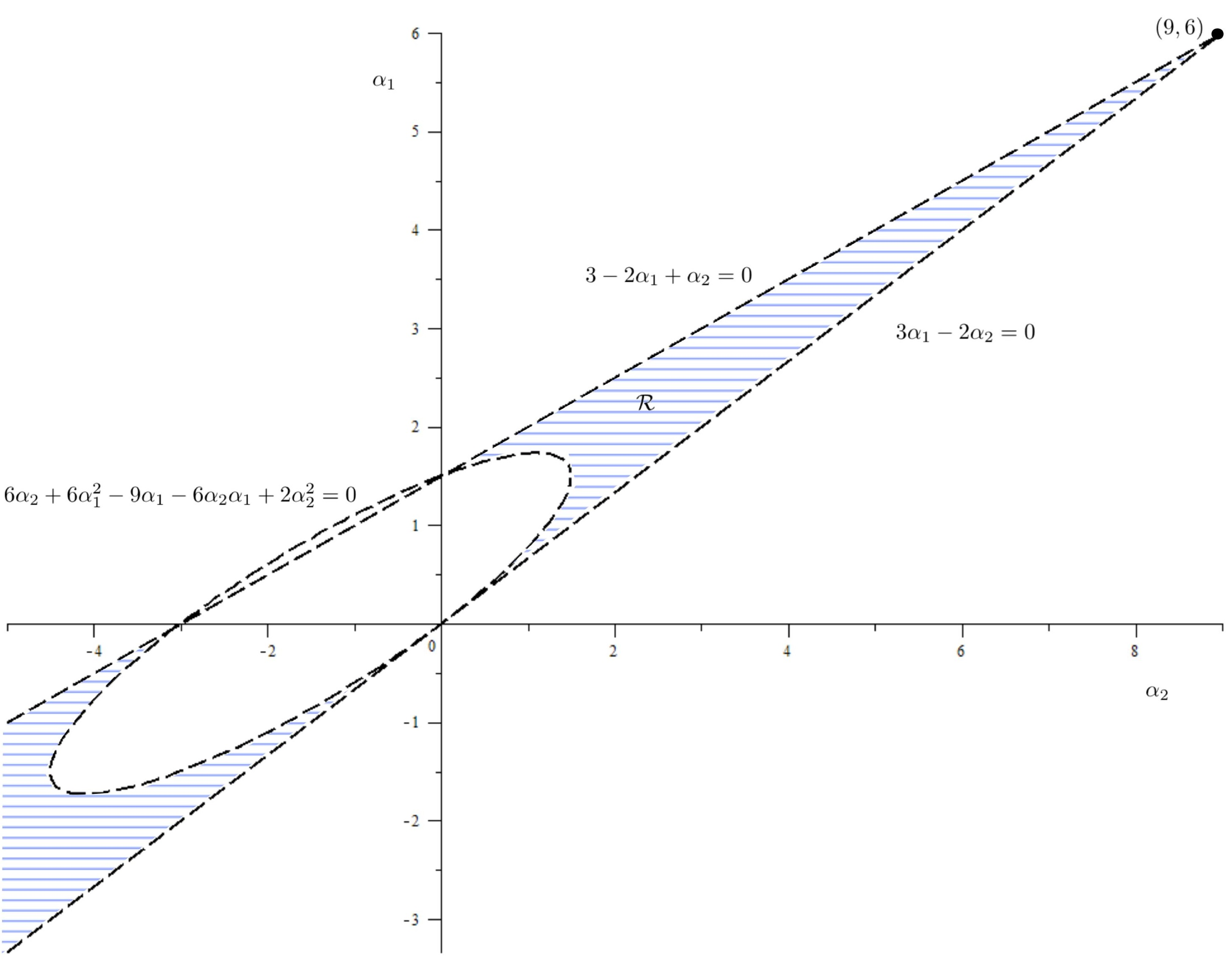}
      \caption{Representation of  the region $\cal{R}$, where the constants $\alpha_1$ and $\alpha_2$ satisfy the strict inequalities
 (\ref{cond1})-(\ref{eq:cond}) and (\ref{cond2}) with $\zeta=1$.}
  \label{fig:figR}
\end{figure}}

Additionally, it is interesting to note that for this new charged black hole, we can analyze its response under electrical fluctuations, represented by the electric permittivity $\epsilon_{T}$ at a constant temperature, which reads as follow:
$$\epsilon_{T}=\left(\frac{\partial \mathcal{Q}_{e}}{\partial \Phi_{e}}\right)_{T}=\frac{72\,\Phi_{e}\kappa^{2}\Omega_{2}}
{\Psi^{2} \zeta^2 {l}^{4}},$$
which is a non-negative quantity {if the strict inequality (\ref{cond2}) holds}, ensuring local stability \cite{Gonzalez:2009nn,Chamblin:1999hg}.

\smallskip{}

%%%%%%%%%%%%%%%%%%%%%%%%%%%%%%%%%%%%%%%%%%%%%
\section{Conclusions and discussions }\label{conclusions}
%%%%%%%%%%%%%%%%%%%%%%%%%%%%%%%%%%%%%%%%%%%%%

In this work we propose a nonlinear electrodynamics in the $(\mathcal{H}, P)$-formalism,
which allows us to obtain charged configurations of AdS black holes in four dimensions with
a {planar base manifold in CG}. These configurations have only one integration constant,
given by the {charge-like parameter} $M$, and {being} parameterized by the structural coupling constants (\ref{eq:Ps}).  \textcolor{black}{As was explained at the beginning, to our knowledge, these planar configurations are the first example of solutions in four-dimensional CG where their thermodynamic quantities do not vanish.}

{With respect to the metric function, the structural coupling constants play a very important role in the characterization of
these charged solutions and when analyzing condition (\ref{eq:condition}), we conclude that there are five
different cases: one represents a black hole with three horizons, two cases represent black
holes with two horizons and the other two are single horizon configurations.}

Also, {in order to} find these new charged black holes, the introduction of nonlinear electrodynamics to CG allows us to obtain \textcolor{black}{nonzero} thermodynamic properties, thanks to the contributions given by $\alpha_1$ and $\alpha_2$. Together with the above,  these configurations satisfy  the four dimensional Smarr relation (\ref{eq:Smarr-4d})  as well as the First Law
(\ref{eq:first-law}). Additionally, the  Critical-Gravity-non-linear electrodynamics model enjoys local stability under thermal fluctuations,
thanks to the non-negativity of the specific heat $C_{\Phi_{e}}$ as well as the Gibbs Free Energy $G$ analysis {if (\ref{cond1})-(\ref{eq:cond}) and (\ref{cond2}) are satisfied}. Supplementing the above, the non-negativity of the electric permittivity $\epsilon_{T}$ shows that our solution is also a locally stable thermodynamic system under electrical fluctuations. \textcolor{black}{It is interesting to note the behavior of $\epsilon_{T}$ for these charged configurations, {as it is a non-negativity quantity if $\Phi_{e}> 0$}, unlike  other solutions found in the literature (see for example \cite{Bravo-Gaete:2021hza}).}

{Some natural extensions of this work may include, the exploration of other gravity theories with quadratic contributions. In this sense, a theory that also showcases critical conditions, is given in \cite{Edery:2019bsh} where the square of the Weyl tensor and the square of the Ricci scalar play the main roles in the action, in the absence of the Einstein Gravity. Another interesting scenario would be to study the higher dimensional case \cite{Alvarez}, where now the CG Lagrangian takes the form
\begin{eqnarray*}
\mathcal{L}&=&R-2\Lambda+\beta_1{R}^2
+\beta_2{R}_{\alpha\beta}{R}^{\alpha\beta}\nonumber\\
&+&\beta_3 {R}_{\alpha \beta \mu \nu} R^{\alpha \beta \mu \nu},
\end{eqnarray*}
and that the coupling constants are tied as  \cite{Deser:2011xc}
\begin{eqnarray*}
\beta_1 =-\frac{\beta_2}{2 (D-1)}=\frac{2 \beta_3}{(D-1)(D-2)}= \frac{1}{4 \Lambda (D-3)},
\end{eqnarray*}
where the four dimensional case (\ref{eq:Squad})-(\ref{relations}) can be recovered impossing $D=4$ together with the transformation
$$(\beta_1,\beta_2,\beta_3) \mapsto (\beta_1-\beta_3,\beta_2+4\beta_3,0).$$}

Given the power of the non-linear electrodynamics as a matter source to find new solutions with a planar base manifold, it would be interesting to {study} charged black holes where their event horizons enjoy \textcolor{black}{spherical or hyperbolical} topologies. \textcolor{black}{It would also be interesting to study charged black hole configurations with non-standard asymptotically behaviors, such as Lifshitz black holes, which were first explored for the uncharged case with CG in \cite{Bravo-Gaete:2021kgt}. For spherically symmetric metrics, it is possible, as was shown in \cite{Karch:2015rpa}, to obtain a generalization of the Smarr relation as well as the first law of black hole mechanics, being understood from a dual holographic point of view and related to the black hole chemistry \cite{Kubiznak:2014zwa,Kubiznak:2016qmn,Sinamuli:2017rhp},  where now the cosmological constant $\Lambda$ takes the role as a dynamical variable, unlike the expression found in (\ref{eq:Smarr-4d}) where we explored  charged planar black holes and $\Lambda$ does not appear in an active way.}

\textcolor{black}{Finally, from a physical motivation, these nonzero extensive thermodynamical quantities will allow us to explore, from a holographic point of view, the connection between black holes and quantum complexity \cite{Brown:2015bva,Brown:2015lvg},  as well as  the effects on shear viscosity \cite{Kovtun:2003wp,Kovtun:2004de,Son:2002sd}, where the mass $\mathcal{M}$ and the entropy $\mathcal{S}_{W}$ take a providential role.}

%%%%%%%%%%%%%%%%%%%%%%%%%%%%%%%%%%%%%%%%%
\section*{Acknowledgments}
%%%%%%%%%%%%%%%%%%%%%%%%%%%%%%%%%%%%%%%%%

The authors would like to thank
Daniel Higuita, Julio M\'endez, Eloy Ay\'on-Beato and Julio Oliva for useful discussion and comments on this work. \textcolor{black}{The authors
thank the Referee for the commentaries and suggestions
to improve the paper.}

%%%%%%%%%%%%%%%%%%%%%%%%%%%%%%%%%%%%%%%%%
\appendix
%%%%%%%%%%%%%%%%%%%%%%%%%%%%%%%%%%%%%%%%%

%%%%%%%%%%%%%%%%%%%%%%%%%%%%%%%%%%%%%%%%%
\section{Analysis of extrema of the black hole solutions}\label{appendix}
%%%%%%%%%%%%%%%%%%%%%%%%%%%%%%%%%%%%%%%%%

In this work, we find that \textcolor{black}{CG} admits black hole solutions provided the existence of a non-linear electrodynamics described by eq. (\ref{eq:H}), and that these solutions are characterized by the gravitational potential (\ref{eq:f}), {where, in our context, $r > 0$}. However, the nature of these solutions and, in particular, the number of horizons that they will exhibit, will depend on the signs and relations between the constants $\alpha_i$'s. In this appendix, we make an analysis of the extrema of the function $f(r)$ to give a general classification of the solutions.
%We begin by determining the extreme values of $f(r)$. First, we calculate the first derivative:
%\begin{equation}
%f'(r)=\alpha_1\sqrt{M}\frac{l}{r^2}-2\alpha_2M\frac{l^2}{r^3}+3\alpha_3M^{3/2}\frac{l^3}{r^4},
%\end{equation}
%and find that the extreme values are located at
%\begin{eqnarray}
%r_{ext1}&=&{\frac { \sqrt{M}\,\left( { \alpha_2}+\sqrt {{{ \alpha_2}}^{2} -3\,{ \alpha_1}\,{
%\alpha_3}}\right) l}{{ \alpha_1}}
%},\label{eq:rext1}\\
%r_{ext2}&=&{\frac { \sqrt{M}\,\left( { \alpha_2}-\sqrt {{{ \alpha_2}}^{2} -3\,{ \alpha_1}\,{
%\alpha_3}}\right) l}{{ \alpha_1}}
%}. \label{eq:rext2}
%\end{eqnarray}
%The next thing that we must do is evaluate this extreme values at the second derivative
%\begin{equation*}
%f''(r)=-{\frac {2 { \alpha_1}\,\sqrt {M}l}{{r}^{3}}}+{\frac {6 { \alpha_2
%}\,M{l}^{2}}{{r}^{4}}}-{\frac {12 { \alpha_3}\,{M}^{3/2}{l}^{3}}{{r}
%^{5}}}
%\end{equation*}
% to determine whether the function $f(r)$ exhibits a maximum or a minimum. This substitution yields to the expressions:
%5\begin{eqnarray}
%f''(r_{ext1})=\frac{
%2\alpha_1^4\sqrt{\alpha_2^2-3\alpha_1 \alpha_3}}{(\sqrt{\alpha_2^2-3\alpha_1\alpha_3}+\alpha_2)^4 Ml^2}, \label{eq:Frr_e1}\\
%f''(r_{ext2})=-\frac{
%2\alpha_1^4\sqrt{\alpha_2^2-3\alpha_1 \alpha_3}}{(\sqrt{\alpha_2^2-3\alpha_1\alpha_3}-\alpha_2)^4 Ml^2}\label{eq:Frr_e2}.
%\end{eqnarray}
Let us recall, from \textcolor{black}{expression (\ref{eq:rext})}, that the existence of extreme values is limited to the fulfillment of the condition (\ref{eq:condition}).
%This will be the starting point, to analyze each case independently.
%\subsection{Arbitrary values of $\alpha_2$}
Let us notice that when $\alpha_1$ and $\alpha_3$ have opposite signs, the condition above is met immediately, regardless of the value or sign of $\alpha_2$. That is, the existence of extreme values is assured. Let us then, start by analyzing both cases in detail.
%%%%%%%%%%%%%%%%%%%%%%%%%%%%%%%%%%%%%%%%%
\subsection{Case $\alpha_1<0,\, \alpha_3>0$}\label{appendix1}
%%%%%%%%%%%%%%%%%%%%%%%%%%%%%%%%%%%%%%%%%
%When analyzing expressions, we notice that the condition (\ref{eq:condition}) is always met for the intervals of interest of $\alpha_1$ and $\alpha_3$, regardless of the sign of $\alpha_2$. That is, there will always be extreme values.
Upon \textcolor{black}{inspecting eqn. (\ref{eq:Frr_e})} when considering $\alpha_1<0,\, \alpha_3>0$, we notice that $r_{ext1}$ corresponds to a minimum and $r_{ext2}$ corresponds to a maximum. Moreover, \textcolor{black}{from (\ref{eq:rext})} , we notice that for these values of $\alpha_1<0$ and $\alpha_3>0$, the maximum will always be on the interval $r>0$ (while the minimum will be in the region $r<0$). Combining this information with the asymptotical behavior of $f(r)$, we can conclude that for the case $\alpha_1<0,\, \alpha_3>0$, the black hole will always have one horizon, as seen in Figure \ref{fig:onehorizon}.
%%%%%%%%%%%%%%%%%%%%%%%%%%%%%%%%%%%%%%%%%
\subsection{Case $\alpha_1>0, \alpha_3<0$}\label{appendix2}
%%%%%%%%%%%%%%%%%%%%%%%%%%%%%%%%%%%%%%%%%

Likewise, the condition (\ref{eq:condition}) is always met when $\alpha_1>0$ and $\alpha_3<0$, regardless of the sign of $\alpha_2$. This means that, in this case too, there will always be extreme values. Again, studying the second derivative of $f(r)$ evaluated in the extreme values (\textcolor{black}{see eqn. (\ref{eq:Frr_e}))}, we notice that $r_{ext1}$ and $r_{ext2}$ correspond to a minimum and a maximum respectively, and that for these values of $\alpha_i$, the minimum will always be on the interval $r>0$ and the maximum in the region $r<0$ ({under our analysis, we suppose that $r > 0$}). If we additionally consider the asymptotical behavior of $f(r)$ (which states that $f(r)$ will be decreasing for small positive values of $r$), we find that for this case the black hole will have a minimum. This means that $f(r)$ can display up to two horizons,
as seen in Figure \ref{fig:twohorizons} Top {black curve}.

For this solution to display both horizons, the strict inequality in (\ref{eq:criticalbh}) must be met, while the strict equality would correspond to {the extremal} black hole ({Figure \ref{fig:twohorizons} Top {red curve}}). On the contrary, if the condition (\ref{eq:criticalbh}) is not met, there will be no horizon and $f(r)$ will not represent the gravitational potential of a black hole.

%\subsection{Specific conditions for all $\alpha_i$}
%%%%%%%%%%%%%%%%%%%%%%%%%%%%%%%%%%%%%%%%%
Having established the configurations that arise when $\alpha_1\alpha_3<0$
 %(and thus, the condition (\ref{eq:condition} is automatically met)
 , let us now analyze the case in which $\alpha_1$ and $\alpha_3$ have the same sign.

\subsection{Case $\alpha_3>0, \alpha_1>0, \alpha_2>0$}\label{app:3horizons}
%%%%%%%%%%%%%%%%%%%%%%%%%%%%%%%%%%%%%%%%%

When analyzing \textcolor{black}{the expression (\ref{eq:Frr_e})} that encode the concavity at the extreme values, we notice that the condition (\ref{eq:condition}) is not always met for the intervals of interest of $\alpha_1$ and $\alpha_2$ and $\alpha_3$. Therefore, it is important to make a separate analysis.
If the condition (\ref{eq:condition}) is met we notice that $r_{ext1}$ corresponds to a minimum and $r_{ext2}$ corresponds to a maximum. Moreover, we notice that for these values of $\alpha_i$, both extrema will always be on the interval $r>0$ (the maximum followed by the minimum). This will result in having three horizons (see Fig. \ref{fig:threehorizons} Bottom). A natural question can be wether there can be a combination of $\alpha_i$ such that the maximum or the minimum coincides with one of the horizons, thus resulting in having only two horizons in total.
In order to obtain an affirmative answer to that question, one must impose that one of the following conditions is met:
\begin{eqnarray*}
0&=&1-\alpha_1\sqrt{M}{\frac {{l}}{r_{ext1}}}+\alpha_2 M {\frac {{l^2}}{{r_{ext1}}^{2}}}-\alpha_3 M^{3/2} {\frac {{l^3}}{{r_{ext1}}^{3}}},\\
0&=&1-\alpha_1\sqrt{M}{\frac {{l}}{r_{ext2}}}+\alpha_2 M {\frac {{l^2}}{{r_{ext2}}^{2}}}-\alpha_3 M^{3/2} {\frac {{l^3}}{{r_{ext2}}^{3}}},\\
\end{eqnarray*}
which is equivalent to imposing:
\begin{equation}\label{eq:degeneration}
\alpha_3=-\frac{\alpha_1^2\alpha_2+2\alpha_1^2 \eta_{\pm}-6 \alpha_2^2-6\alpha_2 \eta_{\pm}}{9\alpha_1},
\end{equation}
with $\eta_{\pm}=\frac{\alpha_1^2}{3}-\alpha_2 \pm \frac{1}{3}\sqrt{\alpha_1^4-3\alpha_1^2\alpha_3^2}$.

\begin{figure}[h!]
\centering
\begin{minipage}[c]{7cm}
\includegraphics[width=7cm]{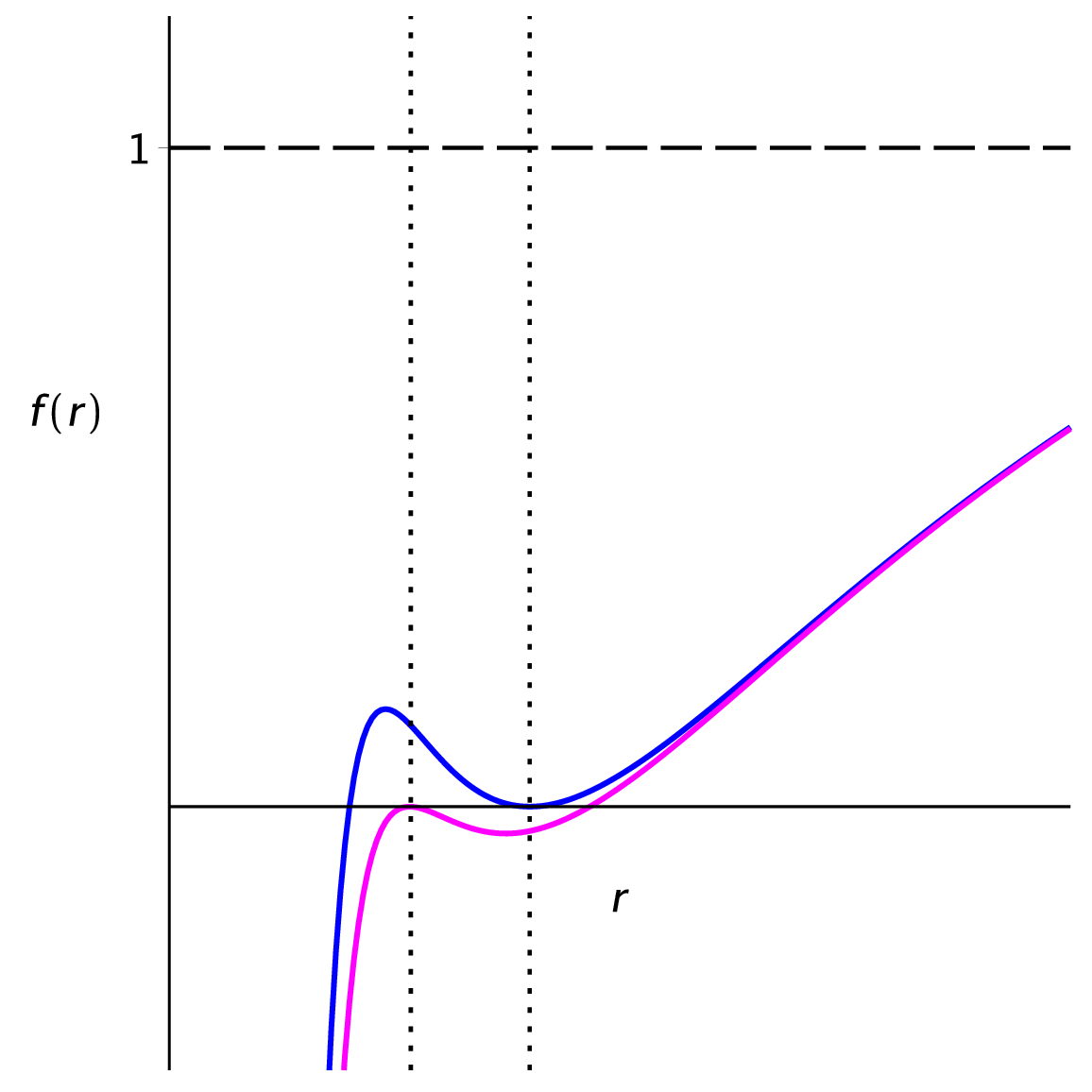}
\end{minipage}
\caption{Gravitational potential $f(r)$ associated to black holes when $\alpha_1>0, \alpha_2>0, \alpha_3>0$ with two horizons when the conditions $3\alpha_1\alpha_3-\alpha_2^2 > 0$ and (\ref{eq:degeneration}) are met.}\label{fig:threehorizons_critical}
\end{figure}

Choosing $\eta_{+}$, in condition (\ref{eq:degeneration}), would correspond to the case in which the minimum coincides with the outer horizon, while choosing $\eta_{-}$ would correspond to the case in which the maximum coincides with the inner horizon. Both cases correspond to solutions with two horizons and are displayed in Figure \ref{fig:threehorizons_critical}.

On the contrary, if the condition (\ref{eq:condition}) is not met, then $f(r)$ will not display extreme values. Due to the asymptotic behavior of $f(r)$ (which states that for small positive values of $r$, $f(r)$ will be increasing, and that it will approach one asymptotically as $r\longrightarrow +\infty$), $f(r)$ will represent the gravitational potential of a black hole with a single horizon as seen in Fig. \ref{fig:threehorizons} Top.

%%%%%%%%%%%%%%%%%%%%%%%%%%%%%%%%%%%%%%%%
\subsection{Case $\alpha_3>0, \alpha_1>0, \alpha_2<0$}
%%%%%%%%%%%%%%%%%%%%%%%%%%%%%%%%%%%%%%%%%

\textcolor{black}{In a similar way,} the condition (\ref{eq:condition}) is not necessarily met when $\alpha_3>0, \alpha_1>0, \alpha_2<0$.
%Therefore, it is important to make a separate analysis.
When the choice of the $\alpha_i$ allows the condition (\ref{eq:condition}) to be met, then $f(r)$ displays a minimum at $r_{ext1}<0$ and a maximum $r_{ext2}<0$ corresponds to a maximum. That is, for these values of $\alpha_i$, both extrema are on the interval $r<0$ of no physical significance. However, we can still obtain important information from the asymptotic behaviour of $f(r)$. As mentioned above, the signs of $\alpha_1$ and $\alpha_3$ will imply that the function $f(r)$, in the interval $r>0$ will start increasing from $-\infty$ and approach one asymptotically from below, representing a black hole with a single horizon (see Fig. \ref{fig:threehorizons} Top).

On the other hand, if the condition (\ref{eq:condition}) is not met, $f(r)$ will not display extreme values at all. Since the asymptotic behavior of $f(r)$ is the same as above, this configuration will also represent a black hole with a single horizon with the shape seen on Fig. \ref{fig:threehorizons} Top.\\

%%%%%%%%%%%%%%%%%%%%%%%%%%%%%%%%%%%%%%%%%
\subsection{Case $\alpha_1<0, \alpha_3<0, \alpha_2<0$}
%%%%%%%%%%%%%%%%%%%%%%%%%%%%%%%%%%%%%%%%%
Let us now consider the scenario in which all the $\alpha_i$ are negative, which means that the condition (\ref{eq:condition}) may or may not be met. As such, it is important to make a separate analysis, as in the previous cases. If the condition, is met we notice that $r_{ext1}$ corresponds to a minimum and $r_{ext2}$ corresponds to a maximum. Moreover, we notice that for these values of $\alpha_i$, both extrema will always be on the interval $r>0$ with the minimum followed by the maximum. A quick analysis of the asymptotic behavior shows that, even though we have more extrema than in previous cases, the maximum number of horizons will be two. The reason for this is that, while the function $f(r)$ starts decreasing in $r>0$, then showcases a minimum and then a maximum, there is not additional crossing of the horizontal axis (since $f(r)$ will approach one from above as $r$ approaches infinity). For clarity, see Fig. \ref{fig:twohorizons} Bottom.

Additionally, we can determine when this solution will display two, one or no horizons through the inequality (\ref{eq:criticalbh}). If the strict inequality is met, {$f(r)$} will represent the gravitational potential of a black hole with two horizons ({Fig. \ref{fig:twohorizons} Bottom black curve}), while the case in which the equality is met strictly would correspond to {the extremal case (Fig. \ref{fig:twohorizons} Bottom red curve)}. On the contrary, if the condition (\ref{eq:criticalbh}) is not met, a horizon will not be formed and $f(r)$ will not be associated to a black hole configuration.

Lastly, if the condition (\ref{eq:condition}) is not met, then, $f(r)$ will not display extreme values. Due to the previously mentioned asymptotic behavior of $f(r)$,  there will be no {$r_h>0$ such that $f(r_h)=0$}. As a result, this case will not correspond to a black hole solution either.

%%%%%%%%%%%%%%%%%%%%%%%%%%%%%%%%%%%%%%%%%
\subsection{Case $\alpha_1<0, \alpha_3<0, \alpha_2>0$ (no solutions)}
%%%%%%%%%%%%%%%%%%%%%%%%%%%%%%%%%%%%%%%%%
Finally,
%when analyzing expressions, we notice that the condition (\ref{eq:condition}) is not always met for the intervals of interest of $\alpha_1$ and $\alpha_2$ and $\alpha_3$. Therefore, it is important to make a separate analysis.
{if} the condition (\ref{eq:condition}) is met and $\alpha_1<0, \alpha_3<0, \alpha_2>0$, the analysis of \textcolor{black}{expressions (\ref{eq:rext})-(\ref{eq:Frr_e})} yields to $r_{ext1}$ being a minimum and $r_{ext2}$ corresponding to a maximum. However, this same set of equations shows that both extrema will be on the interval $r<0$ of no physical significance (that is,  $f(r)$ will not display extreme values in the interval $r>0$). Furthermore, the asymptotic behavior of $f(r)$ shows that, for small positive values of $r$, $f(r)$ decreases from infinity and eventually approaches one from above as $r$ approaches infinity. This asymptotic behavior implies that, unless there are maxima and minima in between these regions of $r$, the function $f(r)$ will not cross the horizontal axis in $r>0$. Since we have established that all the extreme values are in the region $r<0$, { there is no $r_h>0$ such that $f(r_h)=0$}  (all intersections will occur at $r<0$). As a result, this case will not showcase any horizons and thus, does not correspond to a black hole solution.

%%%%%%%%%%%%%%%%%%%%%%%%%%%

\end{document}